# Quantifying Wetting Dynamics with Triboelectrification

*Xiaolong Zhang[a,b], Michele Scaraggi[c,d,e],\*, Youbin Zheng[a], Xiaojuan Li[a], Yang Wu[a,f], Daoai Wang[a]\*, Daniele Dini[d]\*, Feng Zhou[a]\**

[a]*State Key Laboratory of Solid Lubrication, Lanzhou Institute of Chemical Physics, Chinese Academy of Sciences, Lanzhou 730000, China*

[b]*Hubei Key Laboratory of Hydroelectric Machinery Design & Maintenance, China Three Gorges University, Yichang 443002, China*

[c]*Department of Engineering for Innovation, University of Salento, 73100 Monteroni-Lecce, Italy*

[d]*Department of Mechanical Engineering, Imperial College London, South Kensington Campus, London SW7 2AZ, United Kingdom*

[e]*Istituto Italiano di Tecnologia (IIT), Center for Biomolecular Nanotechnologies, Via Barsanti 14, 73010 Arnesano (Lecce), Italy*

[f] *Qingdao Center of Resource Chemistry and New Materials, Qingdao 266100, China.*

*\*Corresponding Author. Tel: +86-931-4968466*

*E-mail: zhouf@licp.cas.cn; wangda@licp.cas.cn; michele.scaraggi@unisalento.it; d.dini@imperial.ac.uk*

## Abstract:

Wetting is often perceived as an intrinsic surface property of materials, but determining its evolution is complicated by its complex dependence on roughness across the scales. The Wenzel state, where liquids have intimate contact with the rough surfaces, and the Cassie-Baxter (CB) state, where liquids sit onto air pockets formed between asperities, are only two states among the plethora of wetting behaviors. Furthermore, transitions from the CB to the Wenzel state dictate completely different surface performance, such as anti-contamination, anti-icing, drag reduction etc.; however, little is known about how transition occurs during time between the several wetting modes. In this paper, we show that wetting dynamics can be accurately quantified and tracked using solid-liquid triboelectrification. Theoretical underpinning reveals how surface micro-/nano-geometries regulate stability/infiltration, also demonstrating the generality of our theoretical approach in understanding wetting transitions. It can clarify the functioning behavior of materials in real environment.

**Introduction.** Materials with tailored wettability have attracted much attention due to the very high demand for this feature in our daily activities and industrial applications. In particular, superhydrophobic materials, originally inspired by the performance of lotus leaves and other natural systems, have become an important scientific focus in the past thirty years [1-3]. Due to the existence of a stable "air mattress" between water and superhydrophobic materials, such surface typically shows a high water contact angle (>150°) and low hysteresis angle (less than 5°). The air mattress acts as an intermediate layer supporting the droplets and decreasing its adherence, via the Cassie Baxter (CB) state [4], hence conferring the superhydrophobic materials excellent antifouling, ant-icing, self-cleaning, and drag reduction properties [5]. However, the non-wetting CB state is typically metastable, with many factors, such as vibration, evaporation, air diffusion and impact [6] being able to drive the wetting transition from the CB to the Wenzel state; this may lead to significant negative outcomes, including the increase of flow resistance and ice adhesion, the aggregation of marine fouling organisms and contaminants [7], an increased blood or bacteria adhesion. Therefore, a real-time and facile monitoring of surface wettability to predict actual substrate performance as well as its behavior throughout products' lifetime would be a major breakthrough in the technology of materials with designed wettability. Several studies have been devoted to the understanding and in-situ monitoring of the wetting transition [8-12]. Duan and co-workers detected surface wettability transition from the CB to the Wenzel state using confocal microscopy [6]. Contact angle and contact angle hysteresis measurements are also typically adopted (including in this work) to qualitatively observe the occurrence of wetting transitions. Nevertheless, such techniques require the adoption of laboratory equipment unlikely to be portable throughout the whole application lifetime. In this respect, being able to directly quantify the wetting dynamics and related wetting transitions by inferring the wetting state through the direct measurement of a physical marker -intrinsically linked to wetting transitions- would be greatly useful for understanding materials functioning performance.

We note that the solid-liquid triboelectric effect is a type of contact electrification in which certain solids become electrically charged after they come into contact and separate from liquids [13-17]. It is highly dependent on different interfacial characteristics [18], i.e. surface functionalization and micro-/nano-structuring [19-25], and strongly linked to the wettability of the solid surfaces, as shown in the schematic representation of MainFig.1. Transient triboelectricity (approximately-rectangular nano- to micro-current) results from screening the triboelectric (surface) charges onto a conductive backplate; however, periodic loading/unloading contacts are needed to generate wave currents. This can happen through water sloshing or intermittent flow caused by e.g. rain or wave impacting triboelectric active surfaces, which is the current focus of blue energy harvesting research [26-27].

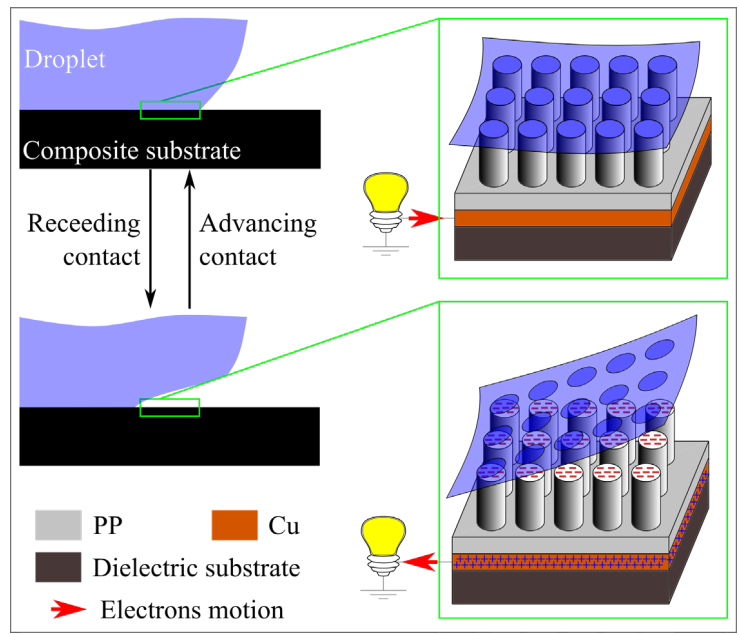

**Figure 1**. Schematic representation of tribocharging and triboelectricity processes, for a generically rough dielectric [polypropylene (PP) in the figure, with copper backplane] in multiple wetting/dewetting contacts with a water droplet. The total amount of tribocharges, thus the tribocurrent, is linearly proportional to the dewetted area during the generic contact cycle.

Differently from the solid-solid contact mechanics and related physics of tribocharge generation, for liquid-based triboelectric nanogenerators (TENGs) no interface elastic energy is stored at the interface during contact, letting the wetting/dewetting dynamics and the total surface charge formation [18, 27, 28] to be regulated only by surface energies. In particular, the detailed roughness across the length scales of the active surface dictates the formation of multiple stable, metastable and unstable liquid-solid contact configurations, depending on the history of fluid squeezing pressure as well as on other environmental conditions [29-32]. Therefore, considering that the amount of true wetting area is intimately linked to the wetting/dewetting and charge generation dynamics [33], thanks to the measurement of the triboelectricity, as we will show below, the dynamic evolution of the wetting state can be quickly detected and reliably quantified, which is otherwise almost impossible to realize.

The research reported in this paper not only sheds light on the mechanisms regulating the intrinsic coupling of triboelectrification and wettability in textured polypropylene surfaces, which can have direct use in the optimization of existing and the development of new TENGs devices, but it also opens a new line of research aimed at exploiting the quantification of such link to quantify dynamic transitions in wetting states through the use of well-calibrated triboelectric systems.

**Results and Discussion.** <u>Micro/nanostructured surfaces</u>. Polypropylene (PP) is selected as the solid of the triboelectric pair, mainly for its low surface energy / high electronegativity and widespread usage in biomedical applications and microfluidics. In our experiments, four typical samples are prototyped, i.e. smooth PP, and PP with random micro-, nano- and hierarchical surface topography, easily replicated by hot press molding (see the schematic of the fabrication process in MainFig.2(A.1) and the Methods for the templates preparation; the FESEM images of the nano-, micro- and hierarchical texture are reported, for different magnifications, in SupFig.SS2B, SS2C and SS2D, respectively). We observe that the choice of random surface textures, instead of deterministically designed textures with predetermined geometries, is due to the simplicity and versatility of molds preparation. The latter requires a maskless lithography process, thus no specific lithography equipment availability and surface size and planarity limitations. This makes them suitable candidates for the facile fabrication of surfaces with tailored and detectable wetting properties. As shown in MainFig.2(A.2), the nano-structured PP tribo-layer is characterized by a forest of nanowires (mean diameter ≈100 nm, SupFig.SS2(B.3)) with similar height (≈1 μm), arranged in randomly distributed clusters as also shown in the virtual prototypes (MainFig.2(A.3-4)). The hierarchical substrate (MainFig.3(A.1)) is characterized by the superposition of the aforementioned nano-pillar forest on the top of a pattern of randomly distributed micro-cubes, see the magnified FESEM image in MainFig.3(A.2 – left). Furthermore, MainFig.3(A.2 – right) shows the micro-structured substrate (without the nanopillars on the top). The PP microcubes have mean side of ≈6 μm.

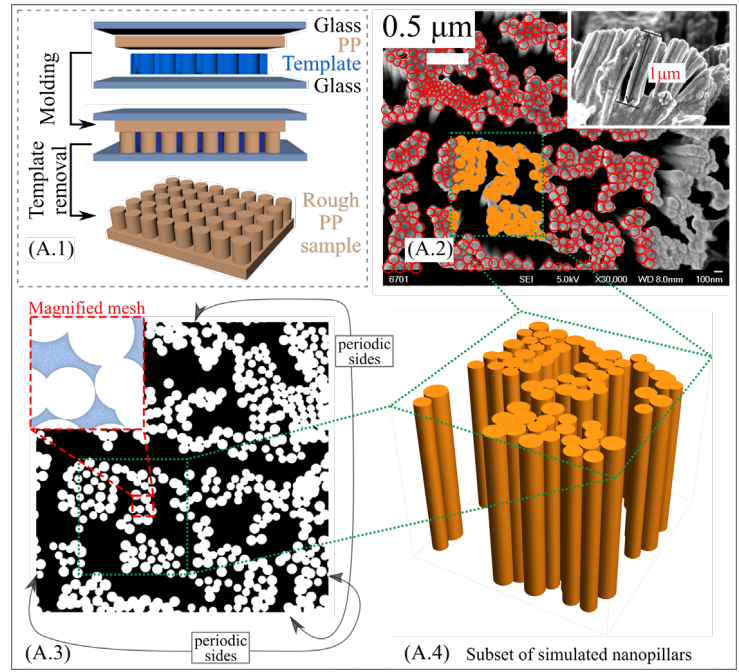

**Figure 2.** Real and virtual prototypes. The different surface roughness obtained on PP substrate after the hot molding process (A.1), including FESEM images of the nanotextured PP (A.2), with corresponding virtual prototypes (A.3 and A.4). All the virtual prototypes have periodicity along the x- and y- direction.

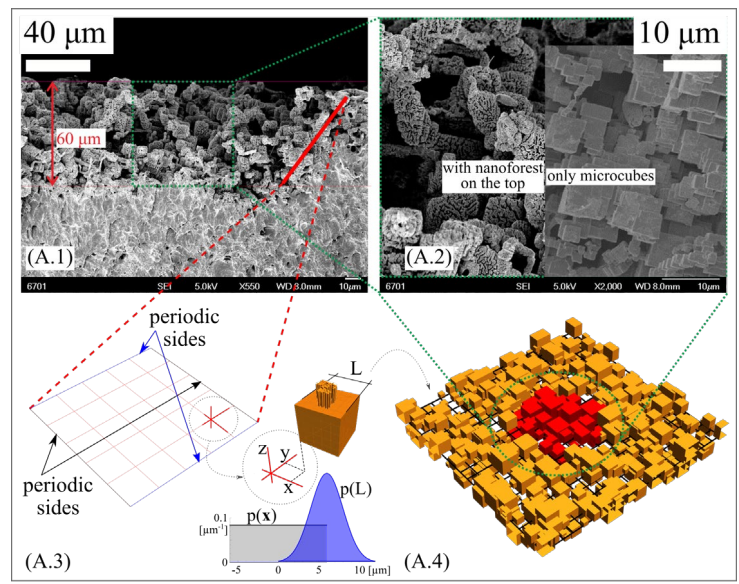

**Figure 3.** Real and virtual prototypes. FESEM image of the random distribution of microcubes (A.1) observed at length scale ≈ 100 μm, and magnified acquisitions of the (A.2-right) microtextured PP and (A.2-left) hierarchical PP, with corresponding virtual prototypes (A.3 and A.4). All the virtual prototypes

have periodicity along the x- and y- direction.

A strong triboelectric signal requires a large amount of reversible surface tribocharges formation, as would be the case when adopting rough surfaces to increase the true interaction area under reversible droplet approach/separation kinematics [34-38]. However, reversibility is highly dependent on the history of wetting pressures of the liquid-surface interactions, thus the triboelectric signal is expected to be strongly dependent on the actual wetting state. To develop models that capture the physical processes linking wetting and tribocharging dynamics, the statistical properties of the fabricated surfaces were extracted from the FESEM images such as reported in MainFigs.2 and 3.

The analysis of a number of surfaces allowed to derive the probability distributions (PDFs, MainFig.3(A.4)) statistically representative of size and position of the micro-blocks, as well as of the nanopillars (SupFig.SS3(A.1)), which were used to generate virtual prototypes (e.g. MainFigs.2(A.4) and 3(A.4)) with the same statistical content of the fabricated surfaces. Note that a relatively large separation of length scales exists between the macroscale system and the microscale texture, as well as between the microscale texture and nanoscale topographic features, which is invoked as we build our multiscale wetting model.

Effective wetting properties through simulations. The contact scenarios between PP and water were simulated by a multiscale approach using the virtual prototypes described in the previous section. The initial step is to derive the surface free energy for a droplet deposited on the different PP surfaces and to study the link between fluid squeezing pressure and wetting states as discussed in Supplementary Materials S1 to S3. Starting from the nano-textured surfaces, it is shown that the pressure needed for a complete transition from Cassie-Baxter (CB) to Wenzel (W) state is relatively large (~MPa) and unlikely to be obtained for the typical operative conditions of the systems under consideration here, leading to a stable droplet true contact area, such as shown in SupFig.SS3(A.3-top). The calculated effective contact angle for the sessile droplet is $\theta_N \approx 123°$. Simulation of micro-textured surfaces in the absence of superimposed nano-texture is done with the numerical model reported in Supplementary Materials S3, using the virtual prototypes as described in MainFig.3(A.4). The droplet-solid separation field can then be computed (see SupFig.SS3(B) and zoomed-in details) and the wetted and free-droplet zones can be determined as a function of the applied fluid pressure, as shown in SupFig.SS3(B.3). At an average squeezing pressure larger than ≈ 200 Pa, an abrupt variation of wetting state occurs (from CB to W), which coincides with a breakdown of the stability of the numerical model. Thanks to the development of the multiscale wetting formulation (Supplementary Materials S1), the application of the nano-pillar structures on top of the micro-textured PP structures (MainFig.4(A.1)) is shown to induce a drastic change of the wetting behavior. The predicted effective contact angle grows drastically with respect to the micro- and nano-textured patterns, and a stable CB state for all pressures relevant to this investigation is found, as shown in MainFig.4(A.2). The hierarchical surface thus behaves as a robust CB-surface, whilst increasing the true contact area with respect to the flat smooth PP thanks to the larger surface area introduced by the micro-texture, see MainFig.3(A.1).

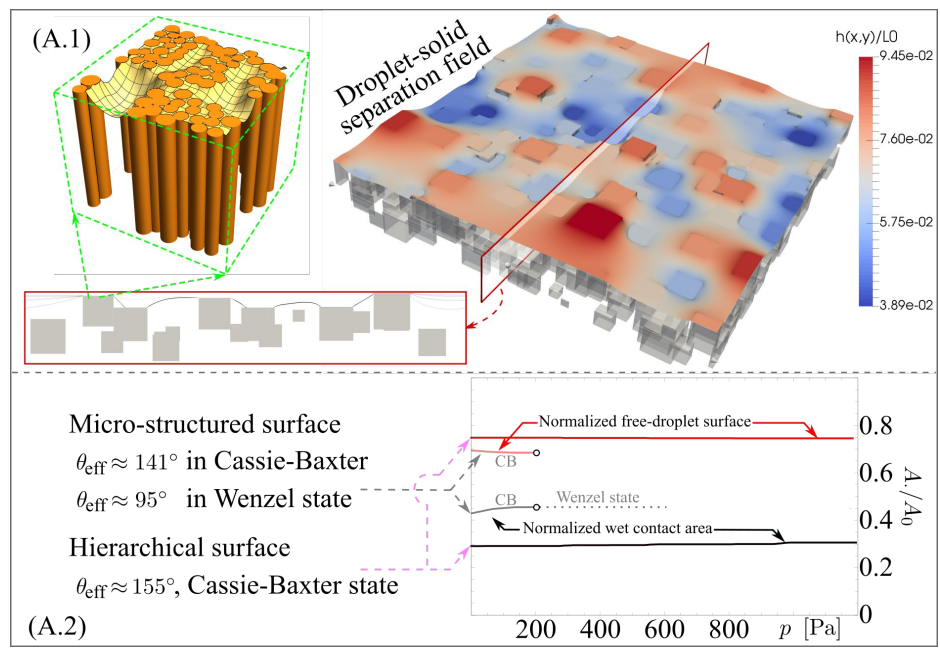

**Figure 4.** Theoretical results. (A.1) Wet contact mechanics from simulations for the hierarchical surface, (A.2) effective contact angle and contact area as a function of the squeezing pressure for the hierarchical and micro-textured surface.

A comparison between the wetting properties predicted by the theory (apparent contact angle, contact radius and contact force) and the results obtained experimentlly are given in SupFig. SS5 for the sessile-droplet experiments, and in MainFigs.5 and 6 for the dynamic squeezing-droplet experiments. For all the shown droplets, the left half represents simulated and right half the real one. It shows the capability of our models to capture the effective or apparent contact angle, thus the true wet contact area, for all investigated surfaces. The dynamic wetting behavior of the four surfaces was investigated since the energy exchanged at the triboelectrification interface, thus the triboelectric signal, is primarily determined by the contact mode at the liquid-solid-gas interface and its evolution at varying droplet squeezing pressures. The sessile droplet was thus pressed and released with a superhydrophobic top-plate (MainFigs. 5 and 6) to shed light on the wetting transition under external pressure, which corresponds to changes of the apparent radius of contact, R, with respect to the initial (sessile) contact radius, $R_0$. Macroscopically, the droplet is shown to undergo different transitions depending on the texture of the PP surface under consideration. For micro-textured PP surface, the droplet does not recover its initial shape after release but shows a much lower contact angle due to the transition from CB to Wenzel state during the compression stage. Data in MainFig.5(A.1 and A.2) shows a marked difference between loading and unloading curves, with a clear transition at around 180 Pa, consistent with the numerical predictions. Interestingly, after the transition to Wenzel state is completed, part of the droplet mass content is trapped in the micro-structured surface, as suggested by the comparison with the predicted sessile droplet shape. On the other hand, the hierarchical PP surface (MainFig.6) shows a remarkably stable CB state. The agreement with the theory in term of droplet pressure vs. displacement curves and apparent contact angles does validate the solid-liquid true contact area theoretical predictions.

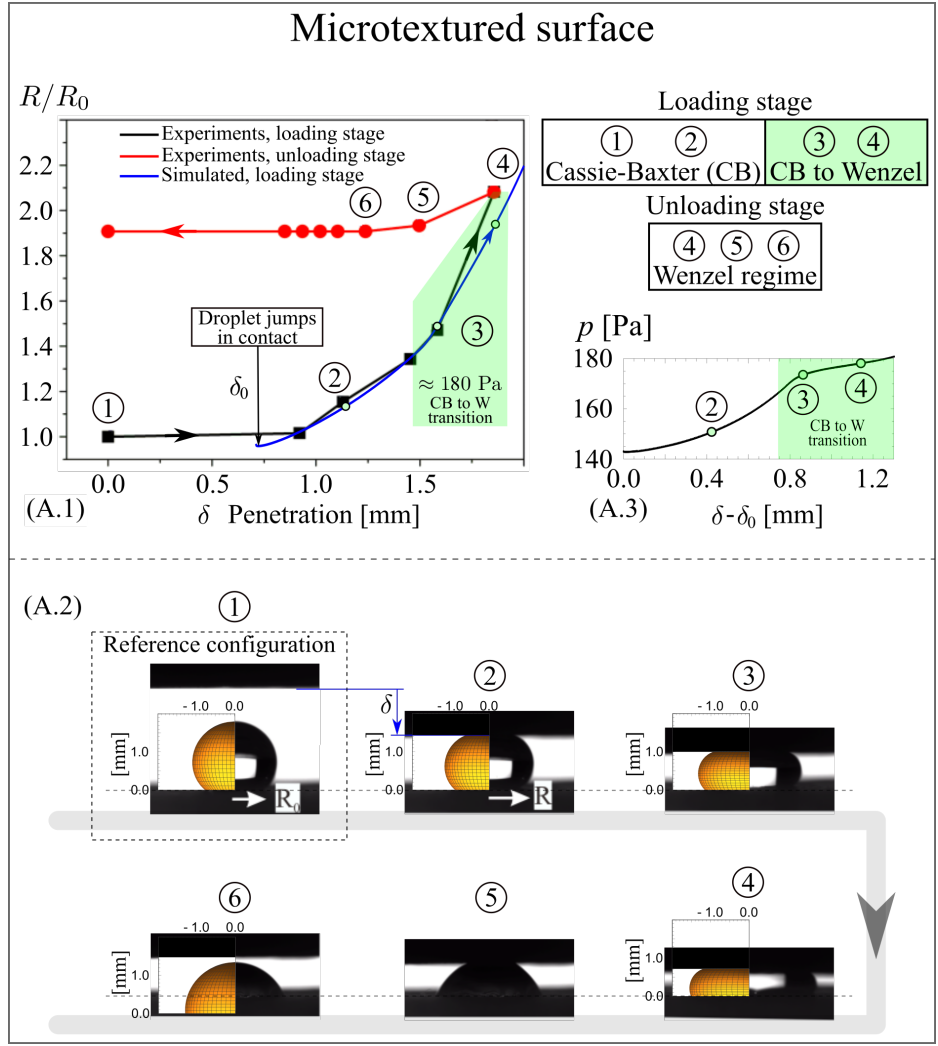

**Figure 5**. Experimental and simulation results of macroscopic dynamic wetting properties, under quasi-static kinematics, for the microtextured surface. In the droplet simulations, the effective contact angle as a function of the droplet relative pressure is taken from the simulations results of MainFig.4 dynamic (quasi-static) wetting properties for the microtextured surface. (A.1) Measured relative contact radius $R/R_0$ as a function of the penetration during loading [(1) to (4), black curve] and unloading [(4) to (6), red curve] stage. The blue curve (only loading stage) is the theoretical prediction, whereas the green area represents the range of pressure and penetration at which the Cassie-Baxter (CB) to Wenzel (W) transition occurs. (A.2) shows the comparison between the predicted and measured droplet shapes at the varying squeezing pressure. (A.3) shows the calculated contact pressure as a function of the penetration. Note the neat curve bending occurring during the CB to W transition, which occurs for a range of water squeezing pressure of about 180 Pa.

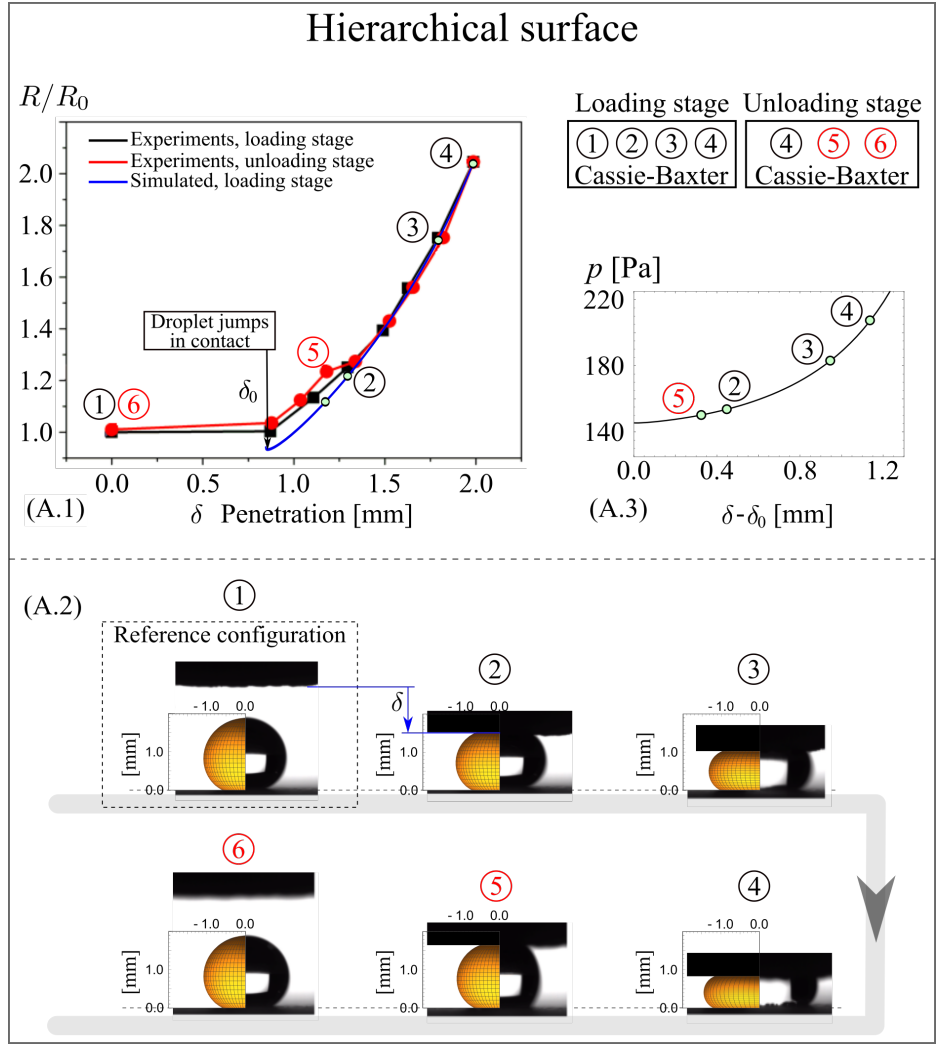

**Figure 6.** Experimental and simulation results of macroscopic static and dynamic wetting properties, under quasi-static kinematics, for the hierarchical surface. In the droplet simulations, the effective contact angle as a function of the droplet relative pressure is taken from the simulations results of MainFig.4 dynamic (quasi-static) wetting properties for the hierarchical surface. (A.1) Measured relative contact radius $R/R_0$ as a function of the penetration during loading [(1) to (4), black curve] and unloading [(4) to (6), red curve] stage. The blue curve (loading/unloading stage) is the theoretical prediction, characterized by a stable Cassie-Baxter (CB) regime. (A.2) shows the comparison between the predicted and measured droplet shapes at the varying squeezing pressure. (A.3) shows the calculated contact pressure as a function of the penetration. $R_0$ is the droplet radius in the sessile contact configuration. Note that the differences between (A.3) and MainFig.5(A.3), in particular for the hierarchical case the curve slope is monotonically increasing with the penetration.

Quantitative measurements of the residual droplet volume with force measurements are given in SupFig.SS8.

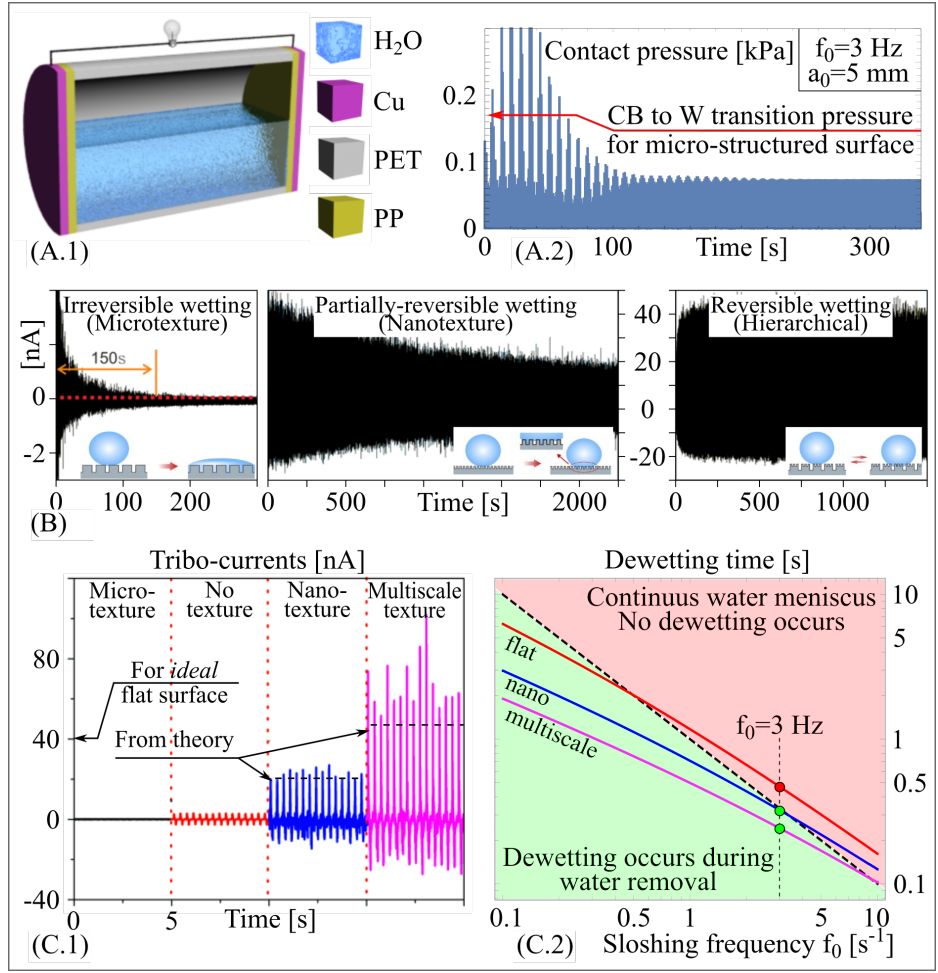

**Figure 7.** Theoretical and experimental results. (A.1) Our TENG device, (A.2) typical TENG sloshing pressure and (C) tribocurrent as a function of time. (D.1) Steady-state tribocurrents and (D.2) drag-out dewetting map.

Quantitative tracing of wetting dynamics by triboelectrification. To experimentally investigate the link between triboelectric signal and dynamic wetting, the PP substrates were integrated onto both ends of a cylindrical tank (schematic in MainFig.7(A.1)), subjected to vibration along the cylinder axis via a linear actuator at varying frequencies. The tank displacement is set sinusoidal, with amplitude of 5 mm. For this prototype macro-geometry, the triboelectrification performance is dictated by the wetting and drag-out dewetting behavior of the surfaces, as well as by the sloshing dynamics associated with the mechanical vibration (MainFig.7(A.2) and SupFig.SS10). The generation of tribo-charges as a function of time for all the structured surfaces developed here is reported in MainFig.7B for the sloshing frequency of 3 Hz, whereas the steady-state tribo-current is reported in MainFig.7(C.1) and compared with the theoretical predictions (see Supplementary Materials S6). The smooth PP film showed relatively small tribocurrent outputs, with a tribo-charge of 0.095 nC (open-circuit voltage Voc of 0.12 V). The PP with micro-textures showed negligible tribocurrent, with a tribo-charge of approximately 0.0029 nC (Voc of 0.0067 V). For the PP with nano-structured surface, the tribo-charge increased to 0.51 nC (Voc of 1.24 V). The hierarchically structured surface produced the highest output with a tribo-charge of about 0.94 nC (Voc of 2.73 V, see Supplementary Materials S7 for more details).

The link between wetting state and tribocurrent dynamics can be fully theoretically explained by investigating the individual phases of the process, starting from the sloshing dynamics induced by the vibration of the water-based TENG constructed here; the key parameters and a schematic representation of the process are reported in SupFig.SS10. The weakly-nonlinear sloshing dynamics model of the partially filled tank (developed as reported in Supplementary Materials S8) enables to predict the moving free surface as well as the center of mass trajectory of the sloshing water for the conditions under consideration and as a function of the filling ratio, as shown in SupFig.SS10(A.2). Capturing the dynamics of the system also allows to determine the average fluid pressure exerted at the wall, and its relation with the shaking frequency; the results reported in MainFig.7(A.2) for example demonstrate how during the transient stage of the sloshing dynamics, the contact pressure exerted on the water surface increases to values which determine the transition from CB to Wenzel wetting state for the micro-textured surface. Thus, for this case, independently of the drag out dynamics, a film of water is cumulatively entrapped onto the PP surface leading to no triboelectric effect, in perfect agreement with the experiments MainFig.7(C.1). However, when the sloshing frequency is reduced, the maximum squeezing pressure encountered during the shaking dynamics decreases to values leading to an incomplete Wenzel transition in the micro-structured PP sample. This is in turn reflected in the measured tribocurrent (see SupFig.SS12), which follows the time dependence of the sloshing pressure (MainFig.7(A.2)). In the weak non-linear sloshing model, the initial increase in the average water pressure is due to the superposition of the steady-state solution, driven by the excitation frequency, and the transient solution, driven by the initial excitation velocity and damped after few cycles. The average water pressure acting on the PP walls is considered in the computation of the tribocharge generation instead of the pointwise pressure function, due to the simplifying assumptions of the sloshing model.

For the smooth, nanotextured and hierarchical surfaces, the triboelectric behavior also depends on the occurrence of dynamic transitions between wetting and drag-out dewetting as a function of the sloshing frequency. The drag-out mechanism can be modelled by generalizing standard drag-out theories as reported in Supplementary Materials S8. The model can be used to describe the dependence of the time to dewet the surface on the spreading pressure and effective contact angle of the different surfaces. This can then be linked to the sloshing frequency to produce the drag-out map reported in MainFig.7(C.2). It is extremely important to note that the triboelectrification is strongly affected by the synchronization between the dewetting time and the sloshing dynamics of the device. MainFig.7(C.2) shows how the different PP surfaces perform in terms of dewetting for difference sloshing frequencies. For the 3 Hz sloshing frequency employed in the experimental set up for demonstration purposes, it is shown how the untextured surface does not reach the post-unstable regime needed for dewetting to take place and for the triboelectric effect to manifest itself. The contrary is true for the nano-textured and hierarchical surfaces, which indeed show transition to dewetting during the sloshing cycle and, thus, a significant triboelectric effect, whose measured values are in agreement with theory. This confirms the intimate link between actual wetting state (and the different physical mechanisms governing this) and nanocurrent signal.

The triboelectric signal originating from the micro-textured surface is remarkably sensitive to its wetting history; in particular, the output current decreases continuously with the sloshing time (SupFig.SS12(A-C)), unravelling the complex wetting behavior of the microstructured PP as made by incremental irreversible local Wenzel transitions. With the increase of the vibration frequency, the time needed for the output to reach the final current also decreases with the different frequencies, from $\approx$ 3000 s at 1 Hz to $\approx$ 150 s at 3 Hz, due to the increase of the squeezing pressure with the sloshing frequency square. On the other side, for hierarchically structured PP (shown for 3 Hz in SupFig.SS12(D)) the tribocurrent is constant, thus the wetted area is unaffected by the variation in squeezing pressure (MainFig.7(A2)), also in agreement with theory.

The theoretical work presented here has enabled us to provide a direct link between the sloshing dynamics and the water pressure building at the interface of the liquid-solid TENG used for the experimental investigation presented in this work. The link between the water pressure applied at the macroscale and the transition between wetting states is indeed the key to unravel the connection between wetting dynamics and triboelectricity; this is confirmed by the fact that the wetting transition responsible for the hysteretic behavior of the droplet squeezed between textured plates, explored in MainFig. 5 and MainFig. 6, can also be directly related to the pressure generated on the liquid film and the wetting transition observed using a sloshing device. This newly gained theoretical understanding has enabled us to link the changes in wettability of the surfaces and their possible reversibility to the physics of the problem and the triboelectric response of the device adopted for this investigation, providing results that are easily transferrable to other scenarios.

It should be highlighted here that the focus of this contribution has been on exploring the effect of texture and surface topography on wetting dynamics. The other important aspect that must be considered to obtain a fuller picture is the effect of surface chemistry and, more specifically, the consequences that chemical changes have on the surface energy and the intrinsic wetting of the surfaces. Although the exploration of this aspect is outside the scope of this work, here we note that the stability of the tribocurrent generation on the hierarchical surface does support the conclusion that the molecular mechanisms allowing triboelectric charging are reversibly occurring on the generic PP surface. However, the robustness of the superhydrophobicity on hierarchical PP can be easily broken, for example by using surfactant solutions, as discussed in detail in Supplementary Materials S10. The presence of surfactant contaminant affects the surface tension of water, which in turn changes the interfacial energy and the equilibrium wetting area. Thus, at increasing surfactant concentration the transition between CB and Wenzel states is favored during the sloshing dynamics, as can be easily detected by the change in the electric signal reported in SupFig.SS13. The higher the surfactant concentration, the faster the decay and the smaller the equilibrium tribocurrent. This confirms the remarkable sensitivity of triboelectricity in the real-time wetting detection and shows the promise that our methodology has in potentially linking chemical changes in both liquid and solid phases to dynamic wetting transitions.

Fluorescence verification. We have employed large field imaging to investigate the link between transient behavior of triboelectricity and fluid infiltration dynamics for the different textured samples, see SupFig.SS15. These observations show that the hierarchical surface displays negligible Wenzel transitions upon multiple sloshing cycles, thus no infiltrated water can be detected on the combined micro/nano-textured surface, as expected theoretically and demonstrated by the intrinsically linked stable triboelectric signal (SupFig.SS12D). On the other hand, the nano-textured PP shows a uniform (due to the sampling resolution of the optical acquisition, which cannot resolve the smallest scales) reduced infiltration after 30 min sloshing time.

This definitely confirms that Wenzel transitions might occur on the nanopillar forest, but without interfering with the dynamics of tribocurrent generation. This agrees with the theoretical study reported in SupFig.SS3A, where we show that Wenzel transitions can occur for the nano-textured surface; however, this does not lead to a substantial variation of the amount of wet area responsible for the triboelectric signal, as this corresponds to the cumulative contribution of all pillars' top surfaces.

As expected, the most interesting infiltration dynamics occurs for the micro-textured surface. In MainFig.8(A.1) the topography and correlation details of the micro-texture surface are reported from a contact stylus acquisition of the surface roughness, shown in magnified view. The largest microcubes edges are clearly visible in terms of highly fragmented small red features in the roughness map. The roughness power spectral density is also reported, with indication of a self-affine domain (with fractal dimension ≈3, related to the random distribution of the microcubes on the surface - see MainFig.3(A.4)) and of the mean microcube frequency $q_{cube} \approx 10^6$ m$^{-1}$, related to an average microcube size $2\pi/q_{cube} \approx 6.3$ μm.

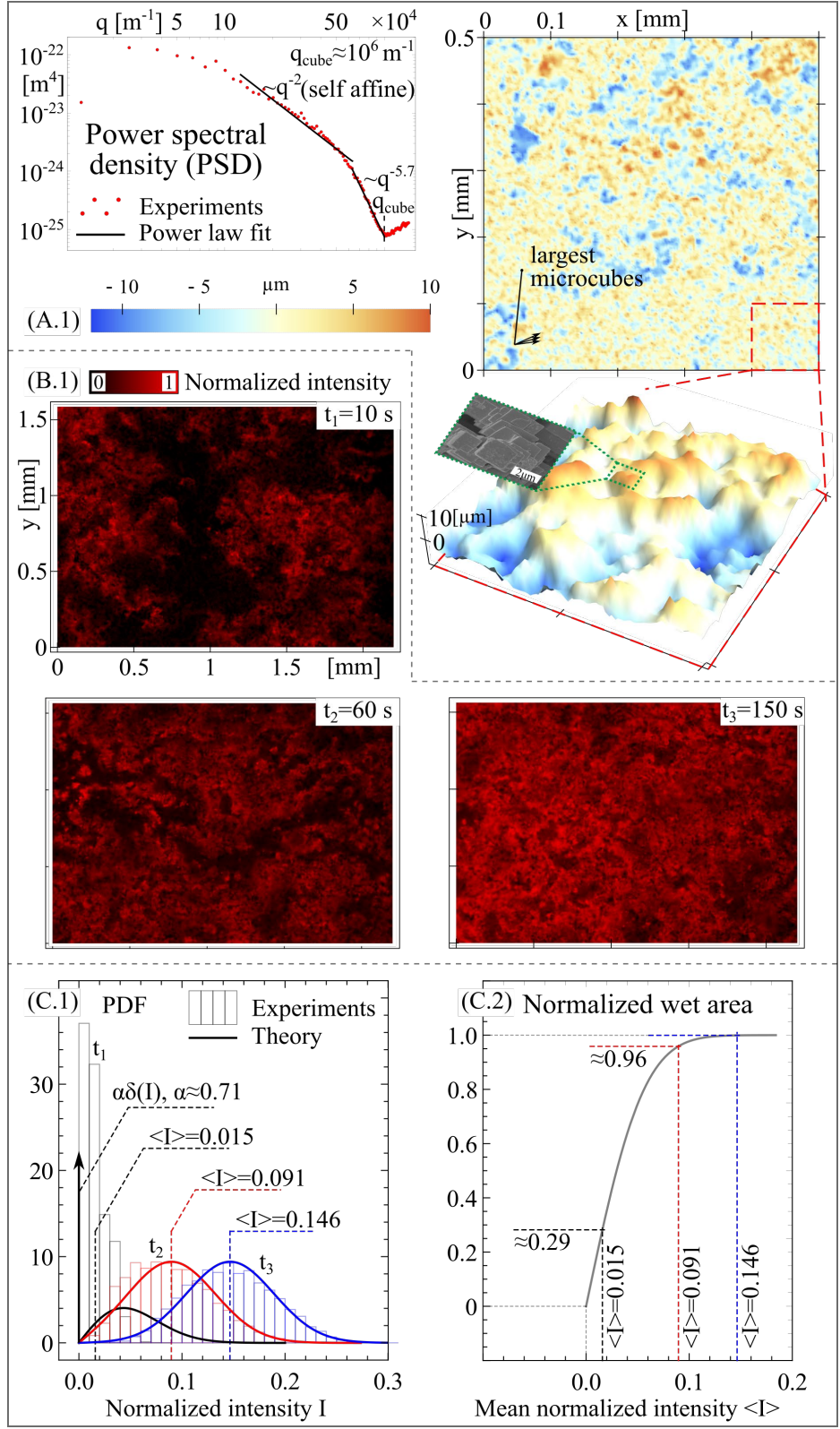

**Figure 8.** Experimental and simulation results. (A.1) Space-of-lengths and spectral description of the microstructured surface. (B.1) Epifluorescence images of the doped water infiltrated in the microtextured surfaces upon multiple sloshing cycles. (C.1) Probability density function of the doped water infiltrated thickness field from experiments and theory ($F(q_1)$ is set to 9E-4 to fit data). (C.2) Evolution of wetted area as a function of the mean infiltrated water volume.

In the panel below, MainFig.8(B.1), the epifluorescence images of the doped water infiltrating the micro-textured surface after multiple sloshing cycles are reported. The black random patterns show, interestingly, that an incomplete irreversible wetting occurs at the water/microtextured PP interface, with the dry domain decreasing at increasing sloshing times. In MainFig.8(C.1) the probability distribution functions (PDF) of the fluorescence intensity are reported against the theoretically predicted PDFs (see Supplementary Materials S11, Equation (S29)), at varying sloshing times. In the theoretical results, we have assumed the fluorescence intensity to be linearly proportional to the wetting film thickness, as applicable in our case. At time $t_1$, the probability is mostly associated to the Dirac's delta function $\delta(I)$ with prefactor $\approx 0.71$ (coarse-grained in the experimental results due to the optics resolution), whereas at time $t_3$ the PDF is mostly Gaussian, in very good agreement with experiments. The theory predicts a PDF of wetting film thickness constituted by the summation of two mirrored Gaussian distributions and a Dirac's delta function (the latter characterizing the probability associated to the non-wetted domains), as a result of a Fokker-Planck equation describing the random film breakdown dynamics, see Supplementary Materials S11. In MainFig.8(C.2) the predicted normalized projected wet area is reported as a function of the average fluorescence intensity (proportional to the average infiltrated water thickness). At time $t_3$ most of the micro-textured surface is wet, with nearly Gaussian distribution of the film thickness, which is again in agreement with the experiments. The irreversibly increasing water infiltration due to a progressive Wenzel transition on the micro-textured surface thus determines the decaying of the triboelectric signal at increasing sloshing times, as clearly shown in SupFig.SS12(C) for the 3Hz case. Remarkably, this demonstrates how triboelectricity can be effectively used for the in-situ monitoring of the most general wetting dynamics on newly designed surfaces, with real-time evaluation of the wet and infiltration contact area according to the theoretical model.

**Conclusions.** The intimate link between triboelectricity and wetting dynamics has been unraveled for (polypropylene) polymer substrates characterized by (i) smooth, (ii) micro-structures, (iii) nano-structures, and (iv) hierarchical structures. A multiscale model was developed to shed light on the mechanisms linking triboelectric generation to wetting dynamics and wetting-induced fluid infiltration. Wetting/dewetting transition from Cassie-Baxter to Wenzel mode, mechanically induced by dynamic water pressure, or from retarded water drag-out process, was correlated with triboelectric decay. This, for the first time, allows us to quantify a plethora of wetting dynamics processes (and related wetting area). On one extreme, the hierarchical PP surface with structured sidewalls provides a stable and large tribo-nanocurrent with respect to untextured and nanostructured surfaces, as a consequence of either a faster drag-out dewetting and a larger reversible true wet contact area. On the other side, incremental infiltration dynamics appears on the microtextured surface, as unraveled too by the triboelectric signal, in real-time. Therefore, the most important outcome of this study is the fidelity of triboelectricity to study the dynamic wetting properties of surfaces. The implications of the results obtained in this work, which also paves the way for designing high-performance liquid-based triboelectric devices, are extremely broad as our findings and the theoretical understanding provided in this paper can be used to e.g. develop design tools and automated testing protocols and sensing technologies to optimize surfaces and tune their wetting dynamics for a plethora of applications for which obtaining and controlling superhydrophilicity is key.

# Experimental section

**Preparation of nanoporous alumina (AA1) templates.**
Aluminum sheets were polished with sandpaper, ultrasonically cleaned in alcohol and washed with deionized water in sequence to get rid of grease. The Al sheet was then polished with an electrochemical method in perchloric acid ($HClO_4$) and ethanol mixture for 10 min with a voltage of 15 V. Finally, a two-steps anodization method was used to prepare the AAO template in 0.3 M oxalic acid ($H_2C_2O_4$) electrolyte with a voltage of 60 V, and the first oxidation process lasted for 2 h whereas the second anodic oxidation time for 20 min. Then, after etching in 5 vol% $H_3PO_4$ solution for 10 min to enlarge the pore size, the nanoporous alumina (AA1) templates were obtained.

**Preparation of micro-structures (AA2) and hierarchical structure alumina (AA3) templates.**
The pre-cleaned Al foil was electrochemically etched in a 10 g/l NaCl aqueous solution at 4 V for 3 h at room temperature to obtain micro-structures (AA2). Secondly, the resultant porous alumina with overhanging micro-structures was anodized at a constant voltage of 60 V for 20 min in 0.3 M oxalic acid at the temperature of 0 °C. Finally, after etching in 5 vol% $H_3PO_4$ solution at 60 °C for 10 min to enlarge the pore size, the hierarchical structure alumina (AA3) templates were obtained. The morphology of AA3 template is given in SupFig.SS16.

**Fabrication of the three different surface roughness PP samples.**
The PP surfaces were prepared by a simplified hot processing technique using AA1, AA2 and AA3 as template, as shown in SupFig.SS2(A). Typically, a flat PP film with the thickness of 50 μm and a template material were squeezed in between of two glass plates, and then moved in oven at 200°C whilst applying a squeezing load of 2 N/cm$^2$. After maintaining the temperature at 200°C for 1 h and cooling down without releasing the load, the prepared PP composite was put into a hot NaOH solution (1.0 M) at 60°C for 3 h to remove the AAO template. Finally, after dissolving the AAO template, the PP friction layers were ultrasonically cleaned in alcohol for 20 min. The obtained structured PP surfaces are shown in SupFig.SS2(B-D).

**Fabrication of triboelectrification device.**
The PP surfaces were used as the solid phase of the triboelectric pair. Copper foil tapes were firstly stuck onto the backside of PP film layer, and then copper wires were attached to one side of the copper tape as a lead wire. Polyethylene terephthalate (PET) cylinder was chosen as the substrate and container material, with cylinder length 200 mm and diameter 40 mm. The PP film was fixed on both ends of the PET tube and sealed with silicone rubber to prevent water leakage. Then, distilled water with/without surfactant was inserted into the tank, acting as the liquid phase of the triboelectric pair, see MainFig.2(B.1).

**Characterization.**
The morphologies of different surface roughness PP were acquired with a field emission scanning electron microscopy (FESEM, JSM-6701F, JEOL Inc., Japan). Contact angle (CA) measurements were done with a DSA100 contact angle meter (Kruss Company, Germany) at room temperature. The average CA value was obtained by measuring the sample at five different positions for 5 μL liquid, and the images were captured with a traditional digital camera. For measurement of the triboelectric outputs of TENG, a sinusoidal reciprocating motion was applied on the tank with a commercial linear mechanical actuator (IVCL17-56) with controllable frequency and amplitude. The open-circuit voltage was measured by using a NI-PCI6259 (National Instruments), while the short circuit current was measured by using an SR570 low-noise current amplifier (Stanford Research System), and data were collected through LabVIEW base Development System (National Instruments)

## Acknowledgements

This research was financially supported by NSFC (51722510, 21773274), Tribology Science Fund of State Key Laboratory of Solid Lubrication (LSL-1903). D.W. would also like to acknowledge the support of the program for Taishan Scholars of Shandong province (No. ts20190965) and the Innovation Leading Talents program of Qingdao (19-3-2-23-zhc) in China. D.D. would also like to acknowledge the support received from the EPSRC under the Established Career Fellowship grant EP/N025954/1. M.S. thanks MIUR for the PRIN 2017 project support under the grant 2017948FEN (FASTire).

## Author contributions

F. Z. and D.W. conceived the idea and supervised the entire research. Y. Z., Y. W. and X. L. performed the experiments and completed the whole characterizations. M. S. and D. D. carried out the simulation. Y. Z. and S. M. drafted the manuscript. F. Z. and D.D. revised and finalized the manuscript. All the authors discussed the results and provided technical suggestions.

## References:

1. L Feng, S Li, Y Li, H Li, L Zhang, J Zhai, Y Song, B Liu, L Jiang, D Zhu Super-hydrophobic surfaces: from natural to artificial. *Adv. Mater.* 14 (24), 1857-1860 (2002)
2. D. Wang, Q. Sun, M. J. Hokkanen, C. Zhang, F. Y. Lin, Q. Liu, S. P. Zhu, T. Zhou, Q. Chang, B. He, Q. Zhou, L. Chen, Z. Wang, R. H. A. Ras, X. Deng, Design of robust superhydrophobic surfaces. *Nature* 582, 55-59 (2020).
3. M. Liu, S. Wang, L. Jiang, Nature-inspired superwettability systems. *Nature Reviews Materials* 2, 1-17 (2017).
4. M. Liu, L. Jiang, Switchable adhesion on liquid/solid interfaces. *Adv. Funct. Mater.* 20, 3753-3764 (2010).
5. B. Su, Y. Tian, L. Jiang, Bioinspired interfaces with superwettability: from materials to chemistry. *J. Am. Chem. Soc.* 138, 1727-1748 (2016).
6. P. Lv, Y. Xue, Y. Shi, H. Lin, H. Duan, Metastable states and wetting transition of submerged superhydrophobic structures. *Physical Review Letters* 112, 196101 (2014).

7. C. Lee, C.-J. Kim, Underwater restoration and retention of gases on superhydrophobic surfaces for drag reduction. *Physical Review Letters* 106, 014502 (2011).
8. Y. Xue, P. Lv, H. Lin, H. Duan, Underwater superhydrophobicity: Stability, design and regulation, and applications. *Applied Mechanics Reviews* 68 (2016).
9. M. Xu, G. Sun, C.-J. Kim, Infinite lifetime of underwater superhydrophobic states. *Physical Review Letters* 113, 136103 (2014).
10. T. Verho et al., Reversible switching between superhydrophobic states on a hierarchically structured surface. *Proc. Nat. Acad. Sci.* 109, 10210-10213 (2012).
11. Y. Li, D. Quéré, C. Lv, Q. Zheng, Monostable superrepellent materials. *Proc. Nat. Acad. Sci.* 114, 3387-3392 (2017).
12. P. Papadopoulos, L. Mammen, X. Deng, D. Vollmer, H.-J. Butt, How superhydrophobicity breaks down. *Proc. Nat. Acad. Sci.* 110, 3254-3258 (2013).
13. Z. H. Lin, G. Cheng, L. Lin, S. Lee, Z. L. Wang, Water–solid surface contact electrification and its use for harvesting liquid-wave energy. *Angew. Chem. Inter. Ed.* 52, 12545-12549 (2013).
14. X. Zhang, Y. Zheng, D. Wang, Z. U. Rahman, F. Zhou, Liquid–solid contact triboelectrification and its use in self-powered nanosensor for detecting organics in water. *Nano Energy* 30, 321-329 (2016).
15. S.-H. Kwon et al., An effective energy harvesting method from a natural water motion active transducer. *Energy & Environmental Science* 7, 3279-3283 (2014).
16. X. Zhang, Y. Zheng, D. Wang, F. Zhou, Solid-liquid triboelectrification in smart U-tube for multifunctional sensors. *Nano Energy* 40, 95-106 (2017).
17. N. Cui et al., Dynamic behavior of the triboelectric charges and structural optimization of the friction layer for a triboelectric nanogenerator. *ACS Nano* 10, 6131-6138 (2016).
18. J.-W. Lee, W. Hwang, Theoretical study of micro/nano roughness effect on water-solid triboelectrification with experimental approach. *Nano Energy* 52, 315-322 (2018).
19. S.-H. Shin et al., Triboelectric charging sequence induced by surface functionalization as a method to fabricate high performance triboelectric generators. *ACS Nano* 9, 4621-4627 (2015).
20. Z. H. Lin et al., A self-powered triboelectric nanosensor for mercury ion detection. *Angew. Chem. Inter. Ed.* 125, 5169-5173 (2013).
21. G. Zhu et al., Toward large-scale energy harvesting by a nanoparticle-enhanced triboelectric nanogenerator. *Nano Letters* 13, 847-853 (2013).
22. F.-R. Fan et al., Transparent triboelectric nanogenerators and self-powered pressure sensors based on micropatterned plastic films. *Nano Letters* 12, 3109-3114 (2012).
23. G. Cheng, Z.-H. Lin, Z.-l. Du, Z. L. Wang, Simultaneously harvesting electrostatic and mechanical energies from flowing water by a hybridized triboelectric nanogenerator. *ACS Nano* 8, 1932-1939 (2014).
24. G. Zhu et al., Harvesting water wave energy by asymmetric screening of electrostatic charges on a nanostructured hydrophobic thin-film surface. *ACS Nano* 8, 6031-6037 (2014).
25. Z. Zhang et al., Oxygen-Rich Polymers as Highly Effective Positive Tribomaterials for Mechanical Energy Harvesting. *ACS Nano* 13, 12787-12797 (2019).
26. D. Jiang et al., Water-solid triboelectric nanogenerators: An alternative means for harvesting hydropower. *Renew. Sust. Energ. Rev.* 115, 109366 (2019).
27. X. Li et al., Networks of High Performance Triboelectric Nanogenerators Based on Liquid–Solid Interface Contact Electrification for Harvesting Low-Frequency Blue Energy. *Adv. Energy Mater*. 8, 1800705 (2018).

28. X. Li et al., Solid–liquid triboelectrification control and antistatic materials design based on interface wettability control. *Adv. Funct. Mater.* 29, 1903587 (2019).
29. H. Cho et al., Toward sustainable output generation of liquid–solid contact triboelectric nanogenerators: The role of hierarchical structures. *Nano energy* 56, 56-64 (2019).
30. T. Vasileiou, J. Gerber, J. Prautzsch, T. M. Schutzius, D. Poulikakos, Superhydrophobicity enhancement through substrate flexibility. *Proc. Nat. Acad. Sci.* 113, 13307-13312 (2016).
31. J. C. Bird, R. Dhiman, H.-M. Kwon, K. K. Varanasi, Reducing the contact time of a bouncing drop. *Nature* 503, 385-388 (2013).
32. Y. Liu et al., Pancake bouncing on superhydrophobic surfaces. *Nature Physics* 10, 515-519 (2014).
33. D. Daniel et al., Mapping micrometer-scale wetting properties of superhydrophobic surfaces. *Proc. Nat. Acad. Sci.* 116, 25008-25012 (2019).
34. S. Lee et al., Triboelectric nanogenerator for harvesting pendulum oscillation energy. *Nano Energy* 2, 1113-1120 (2013).
35. H.-J. Choi et al., High-performance triboelectric nanogenerators with artificially well-tailored interlocked interfaces. *Nano Energy* 27, 595-601 (2016).
36. I.-W. Tcho et al., Surface structural analysis of a friction layer for a triboelectric nanogenerator. *Nano energy* 42, 34-42 (2017).
37. M.-L. Seol et al., Impact of contact pressure on output voltage of triboelectric nanogenerator based on deformation of interfacial structures. *Nano Energy* 17, 63-71 (2015).
38. X. Li et al., Improved triboelectrification effect by bendable and slidable fish-scale-like microstructures. *Nano Energy* 40, 646-654 (2017).

# Supplementary Information

# for

# Quantifying Wetting Dynamics with Triboelectrification

**This PDF file includes:**

    Supplementary text, from Section S1 to S11
    Figs. SS1 to SS16

    SI References

**Supplementary Information Text**

**S1. Effective wetting properties for a multiscale surface**

We consider a representative elementary volume of interface at magnification *n*, where one roughness length scale dominates (say the roughness length scale $\lambda = L_0/n$, where $L_0$ is the system size at the macroscopic scale corresponding to magnification 1). Furthermore, we consider a large separation of length scales between the magnification *n* and the others. Such portion of interface is either in a partial wetted (see Fig.SS1 top left drawing) or dewetted (see Fig.SS1, bottom left drawing) state, depending on the history of the wetting process. In Fig.SS1 we show a schematic (generic) representation of the roughness appearing at magnification *n* and *n+1*, respectively on the left and right panel. In this study we have searched for the equilibrium configurations only for the wetting state, the latter occurring during the initial advancing contact process of the fluid against the rough substrate (see Fig.SS1, top left-right drawing). However, the model below can be similarly applied also to other partial contact states (such as shown in Fig.SS1, bottom left-right drawing).

In particular, we look for the static equilibrium by minimizing the interface energy given by:
$$U = \gamma_{LS}^n A_{0c}^n + \gamma_{SV}^n A_{0s}^n + \gamma_{LV} A_l,$$
where $A_{0c}^n$ is the apparent solid-liquid contact surface, $A_{0s}^n$ the apparent droplet-free surface and $A_l$ the droplet free-surface ($A_0^n = A_{0c}^n + A_{0s}^n$). Above, the superscript *n* refers to the effective physical variables as observed at the actual magnification *n*. By using the definition of effective contact angle observed in the contact at the actual magnification *n*
$$\cos\theta^n = \frac{\gamma_{SV}^n - \gamma_{LS}^n}{\gamma_{LV}}$$
we get
$$U = -\gamma_{LV}\cos\theta^n A_{0c}^n + \gamma_{SV}^n A_0^n + \gamma_{LV} A_l.$$

The equation above can be minimized referring to both the partial wetted (see Fig.SS1, top left drawing) or the dewetted (see Fig.SS1, bottom left drawing) state. However, the effective wetting surface energies $\gamma_{SV}^n$, $\gamma_{LS}^n$ (or $\gamma_{SV}^n$ and $\cos\theta^n$) needs to be calculated from the wetting/dewetting occurring at the scale *n+1*. Thus, when magnifying the domains which appears to be in fluid contact (Fig.SS1, green small boxes on the left panel) or not in contact (Fig.SS1, red small boxes on the left panel) at scale *n*, such domains are fragmented over multiple contacts (Fig.SS1, green large boxes on the right panel) or multiple non-contacts (Fig.SS1, red large boxes on the right panel). By equating the surface energy written in the apparent interaction areas at magnification *n* and *n+1* we get
$$\gamma_{LS}^n = \alpha_{LS}^{n+1}\gamma_{LS}^{n+1} + \alpha_{LV}^{n+1}\gamma_{LV} + \alpha_{SV}^{n+1}\gamma_{SV}^{n+1}$$
and
$$\gamma_{SV}^n = \beta_{LS}^{n+1}\gamma_{LS}^{n+1} + \beta_{LV}^{n+1}\gamma_{LV} + \beta_{SV}^{n+1}\gamma_{SV}^{n+1},$$
where the $\alpha_i^{n+1}$ and $\beta_i^{n+1}$ are the normalized apparent interaction areas between the phases-*i*, respectively, inside and outside the apparent droplet-*n* contact area. In the equations above we have assumed that the effective surface energies occurring, at the generic magnification *n+1*, are equal in both the partial wetted and partial dewetted state, i.e. we have implicitly considered the fluid to be subjected to the same pressure both in the wetting and dewetting areas during the generic interaction process. We observe, however, that this assumption does not affect the calculation in the initial advancing contact process analyzed in this study.

Given that $r_{sub}^{n+1} = \alpha_{SV}^{n+1} + \alpha_{LS}^{n+1} = \beta_{SV}^{n+1} + \beta_{LS}^{n+1}$ is the normalized roughness surface area, one gets the effective equations
$$\gamma_{LS}^n = \gamma_{LV}(\alpha_{LV}^{n+1} - \cos\theta^{n+1}\alpha_{LS}^{n+1}) + r_{sub}^{n+1}\gamma_{SV}^{n+1} \tag{S1}$$
and
$$\gamma_{SV}^n = \gamma_{LV}(\beta_{LV}^{n+1} - \cos\theta^{n+1}\beta_{LS}^{n+1}) + r_{sub}^{n+1}\gamma_{SV}^{n+1} \tag{S2}$$
leading to an effective contact angle
$$\cos\theta^n = \cos\theta^{n+1}(\alpha_{LS}^{n+1} - \beta_{LS}^{n+1}) - (\alpha_{LV}^{n+1} - \beta_{LV}^{n+1}). \tag{S3}$$
Equations (S1-S3) need to be applied recursively for the case of a multiscale roughness, as is the case in our investigation, as shown in Fig.SS2. Clearly, in order to determine the effective wetting properties at scale-*n* ($\gamma_{LS}^n$, $\gamma_{SV}^n$ and $\cos\theta^n$), one needs not only the effective wetting properties at scale *n+1* ($\gamma_{LS}^{n+1}$, $\gamma_{SV}^{n+1}$ and $\cos\theta^{n+1}$), but also the fraction of fluid surface in contact/non-contact with the substrate at magnification *n+1*. For our systems, we have three representative length scales: the macroscale of the droplet (say *n=1*), the intermediate micro-scale (given by the randomly placed micro-cubes, say magnification *n=2*) and the highest magnification at the nano-forest scale (say *n=3*). Thus, we have solved first the static contact process at the nanoscale in Sections S2, whose results are then used to minimize the interface energy in Section S3 at the microscale. Finally, the results from Section S3 are adopted to determine the effective surface energies at the macroscale droplet contact problem.

## S2. Effective wetting properties for the nano-textured surface

Here we study first the effective wetting properties of a hexagonal lattice of cylindrical pillars, with a flat top surface. The pillars have radius $R_p$, height $h_p$ and intra-cluster lattice size $L_p$. We assume the generic local maximum penetration of the droplet in the pillar forest, $\delta$, to be $\delta \ll L_p$ (i.e. local approximation of the droplet surface as paraboloid, valid for small droplet penetration compared to the hexagonal lattice size $L_p$); by applying the Young-Laplace equation, this leads to $\delta \approx (L_p - 2R_p)^2 p/(8\gamma_{LA})$, where $p$ is the droplet relative squeezing pressure and where we have assumed the mean droplet curvature to be $\approx 8\delta/(L_p - 2R_p)^2$. The maximum fluid pressure allowed to keep a Cassie-Baxter (CB) wetting regime, by geometric considerations, is the minimum between

$$\frac{p_{\max} L_p}{\gamma_{LA}} = -\cos\theta_0 \frac{\frac{8R_p}{L_p}}{\frac{2\sqrt{3}}{\pi} - \frac{4R_p^2}{L_p^2}} \tag{S4}$$

and

$$\frac{p_{\max} L_p}{\gamma_{LA}} \approx \frac{h_p}{R_p} \frac{\frac{8R_p}{L_p}}{\left(1 - \frac{2R_p}{L_p}\right)^2}, \tag{S5}$$

whereas the average penetration occurring for pressures $p_0 < p_{\max}$ is:

$$\delta_p \approx \frac{(L_p - 2R_p)^2}{8} \frac{p_0}{\gamma_{LA}} \left(1 - \frac{R_p^2}{L_p^2} \frac{2\pi}{\sqrt{3}}\right).$$

Given that $\frac{8R_p/L_p}{2\sqrt{3}/\pi - 4R_p^2/L_p^2}$ or $\frac{8R_p/L_p}{(1-2R_p/L_p)^2}$ are factors of order 1 for our system, the pressure required for the wetting transition from the CB to Wenzel (W) state will thus be related to the minimum value between the pillar aspect ratio, $h_p R_p^{-1}$ and the effective contact angle, $-\cos\theta_0$. For our nano-forest $2R_p \approx 100$ nm, $h_p \approx 1$μm. Therefore, $h_p R_p^{-1} \approx 20$, whereas $-\cos\theta_0 \approx 0.2$, leading to the conclusion that, for our system, (S4) is the relevant equation describing the squeezing pressure necessary for the transition CB to W to occur for a hexagonal lattice of flat pillars. Eq. (S4) shows that the minimum intra-cluster distance at which the CB to W transition occurs quickly decreases at increasing squeezing pressures. This is shown in Fig.SS3, where it is shown (A3) that a Wenzel transition would occur at intra-cluster distances of about 3 μm for a squeezing pressure of order kPa. This intra-cluster value is much larger than the average intra-cluster distance occurring in our system, see Fig.SS3(A.1 - bottom) and below.

For our system, pillars are arranged in local clusters, with inter-cluster distance probability distribution reported in Fig.SS3(A.1 - top), and intra-cluster lattice spacing probability distribution reported in Fig.SS3(A.1 - bottom). The latter is well fitted with a double Gaussian (see solid line in the above-mentioned figure)

$$P(L_p > L_1) \approx a_0 \left( e^{-\frac{1}{2}\frac{(L_p - L_1)^2}{L_2^2}} - e^{-\frac{1}{2}\frac{(L_p + L_1)^2}{L_2^2}} \right), 0 \text{ elsewhere} \tag{S6}$$

where $\int dL_p\, P(L_p) \approx 0.57$, representing the total top surface pillars area normalized with the substrate area. The CB to W transition is dictated by the wetting occurring at the cluster scale. Thus, in order to predict the effective contact angle for the random nano-pillar forest, we have developed a simple multi-cluster wetting contact model, where clusters of differently spaced hexa-lattice of pillars are randomly distributed accordingly to Eq. (S6), and where one assumes that CB occurs for $L_p < L_{th}(p)$, and the W regime occurs elsewhere. Therefore, once determining for the hexa-lattice the normalized contact areas $A_{LV,()}$, $A_{LS,()}$ and $A_{SV,()}$, where LV, LS and SV is for liquid-air, liquid-solid and solid-air interface, and where () is to indicate the interface in the CB and W contact condition (relations dropped here for simplicity), the total wet contact area can be calculated with

$$\frac{A_{LS}(p_0)}{A_0} = \int_0^{L_{th}(p_0)} dL_p\, P(L_p) A_{LS,CB}(L_p) + \int_{L_{th}(p_0)}^{\infty} dL_p\, P(L_p) A_{LS,W}(L_p)$$

and similarly for the liquid-air and solid-air interfaces. This provides, upon using Eqs. (S1-S3), the wetting predictions reported in Fig.SS3(A3 - top). In particular, for our nanotexture, a nominal CB regime is expected up to pressures much larger than those obtained in the experiments because of the sloshing action, i.e. a complete W transition occurs starting from order $10^5$ Pa pressures.

In order to verify the prediction of the statistical model above, we have performed deterministic numerical resolution of the Young-Laplace differential equation which reads (neglecting high order terms in the local droplet mean curvature radius, which is the case here)

$$\Delta u(\mathbf{x}) - \frac{p_0}{2\gamma_{LV}} = 0, \tag{S7}$$

where $p_0$ is the droplet squeezing pressure and $\gamma_{LV}$ the water surface tension. In the model, the pillars are assumed with same height (consistently with the real nano-pillar forest surface), but randomly distributed on the substrate. Several realizations of the virtual prototypes (both numerically generated or experimentally measured, see e.g. Fig.1(A.2- A4) and Fig.SS3(A.2)) have been simulated. In particular, in Eq. (S7) $u(\mathbf{x})$ is the droplet separation field measured from the (common) pillar top height. Eq. (S7) is thus solved on a periodic forest system (with periodicity boundary conditions, see Fig.1(A.3)), considering $u = 0$ on the pillars (top) edges, adopting mesh refining close to the pillar edges. The results were considered in the CB state if over none of the simulation triple lines the contact angle between pillar and water surface was overcoming the thermodynamic value. The simulations, performed within Mathematica (see an example of results in Fig.SS3(A.2)), confirmed that no CB to W transition was occurring for pressures below $10^5$ Pa order, in agreement with the simple multi-cluster statistical model provided above. At a squeezing pressure of 2 kPa (quite above the maximum sloshing pressure in the experiments, see below), the droplet penetration field for one pillar forest realization is reported in Fig.SS3(A.2) (penetration exaggerated for easy of visualization).

By using Eq. (S3), the effective contact angle provided by the nanotexture is $\approx 123°$, and can thus be considered constant in our experiments. Indeed, considering $\beta_{LS}^N = \beta_{LV}^N = 0$, $\alpha_{LV}^N \approx 0.43$ (measured) and $\alpha_{LS}^N + \alpha_{LV}^N \approx 1$, this leads to an effective contact angle on the lower magnification (say, the microscale)

$$\cos\theta^M = \cos\theta_0 \alpha_{LS}^N - \alpha_{LV}^N$$

obtaining $\theta^M \approx 123°$ and $\gamma_{LS}^M = -\gamma_{LV}\cos\theta^M + \gamma_{SV}^M$, with

$$\gamma_{SV}^M = r_{sub}^N \gamma_{SV},$$

where ($r_{sub}^N \approx 22.8$).

Therefore, if one considers the surface only made by the nano-texture, the total surface energy of a droplet (larger than representative nano-texture size) deposited onto the substrate is

$$U = -\gamma_{LV}\cos\theta^M A_{SL} + r_{sub}^N \gamma_{SV} A_0 + \gamma_{LV} A_{LV}, \tag{S8}$$

with $\theta^M$ independent of the droplet squeezing pressure for pressures lower than about $10^5$ Pa.

## S3. Effective wetting properties for the micro-textured surface

In this section we report the novel computational wetting model for the investigation of a three-dimensional random surface. We stress that the model reported in the following can be applied to any roughness topography, both with random or ordered geometries, and as such is applied to the simulation of the microtextured surface adopted in this research.

As stressed in Section S1, to integrate out the micro-texture contribution to the wetting dynamics when observing the system at lower magnifications, we use the effective equations (S1-S3) over the micro-texture. Here we calculate $\alpha_{LS}^M$ and $\alpha_{LV}^M$, to be used in Eqs. (S1-S3), by studying the wetting contact dynamics occurring on the microtexture. The total (micro-) system energy is

$$\frac{U}{\gamma_{\text{LV}}} = A_{\text{l}} - \cos\theta^{\text{eff}} A_{\text{c}} + \frac{p}{\gamma_{\text{LV}}} u A_0 + \frac{\gamma_{\text{SV}}^{\text{eff}}}{\gamma_{\text{LV}}} A_{\text{tot}} \quad (S9)$$

where $A_l$ is the liquid free surface, $A_c$ is the wet solid surface, $A_{tot}$ is the total solid-phase surface. $\cos\theta^{\text{eff}}$ and $\gamma_{\text{SV}}^{\text{eff}}$ are the contact angle and solid-vapor surface tension, respectively, as oberved on the microtextured surface. Furthermore,

$$dA_{\text{l}} = (1 - f_{\text{c}}(\mathbf{x}))\sqrt{1 + |\nabla u|^2}$$
$$dA_{\text{c}} = f_{\text{c}}(\mathbf{x})\sqrt{1 + |\nabla u|^2}$$

where $u(\mathbf{x})$ is the droplet separation measured from the average roughness plane (i.e. $\langle d(\mathbf{x})\rangle = 0$), and where $f_{\text{c}}(\mathbf{x})$ is the wetting contact index defined as

$$f_{\text{c}}(\mathbf{x}) = e^{-\frac{u(\mathbf{x}) - d(\mathbf{x})}{\rho}},$$

with $d(\mathbf{x})$ the surface roughness and $\rho$ a characteristic length describing the distance over which the fluid-solid contact occurs.

In the case the nanotexture *is* included on the top of the microtexture (see e.g. Fig.1(B.2 - left)), $\cos\theta^{\text{eff}} = \cos\theta^{\text{M}}$ and $\gamma_{\text{SV}}^{\text{eff}} = r_{\text{sub}}^{\text{N}}\gamma_{\text{SV}}$ are provided in Section S2, whereas in the case the nanotexture *is not* included $\cos\theta^{\text{eff}} = \cos\theta_0$ and $\gamma_{\text{SV}}^{\text{eff}} = \gamma_{\text{SV}}$ as given by the pristine PP substrate. Either cases (pristine PP and nanotextured PP) do not show pressure dependence in the terms $\cos\theta^{\text{eff}}$ and $\gamma_{\text{SV}}^{\text{eff}}$ wetting dynamics, therefore Eq. (S9) can be simplified to

$$\frac{U}{\gamma_{\text{LV}}} = A_{\text{l}} - \cos\theta^{\text{eff}} A_{\text{c}} + \frac{p}{\gamma_{\text{LV}}} u A_0. \quad (S10)$$

Eq. (S10) is numerically minimized with the finite element method in the variable $u(\mathbf{x})$ within the FEniCS software environment, where $\mathbf{x} = (x, y)$ is the plane of the nominally flat substrate, under the constraint $u(\mathbf{x}) \geq 0$ (no water intrusion in the solid phase), at constant applied droplet pressure *p*. The computational domain, a square periodic in the *-x* and *-y* directions with periodicity lengths $L_{0x} = L_{0y} = L_0$ (see Fig.1(B.4) and Fig.SS3B), is discretized with an unstructured mesh of Lagrangian triangular elements of order one (see Fig.SS3(B.2)). The mesh is recursively hierarchically refined close to the triple lines and, correspondingly, the total interface free energy minimized at decreasing values of the characteristic length $\rho$. The solver adopted for the non-linear constrained minimization is the TRON algorithm, active-set Newton trust region method for bound-constrained minimization. TRON is made available by the Tao (Toolkit for Advance Optimization) library within the PETSc (Portable, Extensible Toolkit for Scientific Computation) solver in FEniCS (https://fenicsproject.org/).

Once the equilibrium configuration is computed at a given droplet pressure *p*, we determine $\alpha_{\text{LV}}^{\text{M}} = A_{\text{l}}/A_0$ and $\alpha_{\text{LS}}^{\text{M}} = A_{\text{c}}/A_0$ to be used in the effective equations (S1-S3) over the micro-textured surface. The predicted effective wetting parameters, with and without nanoscale texture, are reported in Fig.2(A.2).

In particular, (only) for the case of micro-textured surface a breakdown of the simulations, for the different surface realizations, occurs at squeezing pressures ≈180 Pa (in between ≈150 and ≈200 Pa for the set of run simulations); this suggests that the numerical algorithm is not able to handle the abrupt variation of equilibrium solution across such pressure values, as a consequence of, possibly, a Wenzel transition. We will show in Supplementary Section S4 (see also Fig.SS5 where we show the comparison between theory and experiments) that, very interestingly, a Wenzel transition is exactly the phenomenon occurring when the squeezing pressure achieves such limiting value on micro-textured surface. Furthermore, in order to model the wetting for pressure above the Wenzel transition pressure, we just consider no free-droplet surface in Eq. (S3), leading to a predicted effective contact angle in the Wenzel regime to be ≈95°. The latter result is in very good agreement with experiments, see Fig.SS5. In the CB regime, instead, the calculated effective contact angle for the micro-textured surface is ≈141°, with relatively small dependence with the squeezing pressure (until the W transition), see Fig.SS3B.

Finally, the case where the nano-texture is superposed to the micro-texture (hierarchical surface) shows a stable CB regime in the range of tested squeezed pressures, see Fig.2(A.2). Thus, very interestingly, the addition of nanotexture strongly affects the wetting dynamics occurring on the substrate, leading to a stable effective contact angle of ≈155°, see the simulation results of Fig.2(A.2). Again, this is in very good agreement with the experimental results, see Fig.SS5.

## S4. Theory of droplet statics among two plates

In this section we report the theory developed for the static contact of a droplet among two nominally flat (chemically and physically) dissimilar surfaces. In particular, a droplet with volume $V_0$ and density $\Delta\rho + \rho_A$, where $\rho_A$ is the surrounding out-of-droplet fluid density, is squeezed in between two rigid parallel nominally flat surfaces at distance $h_0$, as schematically reported in Fig.SS6.

The energy $F$ of the system is given by the sum of the potential energy $U_g$ and the surface energy $U_s$, with

$$U_g = \pi \Delta\rho g \int dz\, z r^2(z)$$

and

$$U_s = \gamma_{LA} A_L + \gamma_{LS_1}^{\text{eff}} A_{S_1} + \gamma_{LS_2}^{\text{eff}} A_{S_2} + \gamma_{S_1A}^{\text{eff}} A_{NS_1} + \gamma_{S_2A}^{\text{eff}} A_{NS_2}. \qquad (S11)$$

The mesoscopic mechanical equilibrium requires that

$$\gamma_{LA} = \frac{\gamma_{SA}^{\text{eff}} + \gamma_{LS}^{\text{eff}}}{\cos\theta^{\text{eff}}}, \qquad (S12)$$

where $\theta^{\text{eff}}$ is the effective droplet contact angle measured on the generic nominally smooth substrate. Thus, we write Eq. (S11) with (S12)

$$U_s = \gamma_{LA}\left[A_L - \cos\theta_1^{\text{eff}} A_{S_1} - \cos\theta_2^{\text{eff}} A_{S_2}\right] + U_{s0}^{\text{eff}},$$

with $U_{s0}^{\text{eff}} = \gamma_{S_1A}^{\text{eff}} A_1 + \gamma_{S_2A}^{\text{eff}} A_2$, where $A_1$ ($A_2$) is the bottom (top) apparent substrate area. Furthermore, we have that

$$A_{S_1} = \pi \int dz\, r^2(z) \delta(z)$$

$$A_{S_2} = \pi \int dz\, r^2(z) \delta(z - h_0),$$

where $h_0$ is the distance between the plates (and $\delta(z)$ is Dirac delta), and where

$$A_L = 2\pi \int dz\, r \sqrt{1 + r_{,z}^2}.$$

We make all lengths **dimensionless** with $l_0 = V_0^{1/3}$, and energies with $\gamma_{LA} l_0^2$, leading to

$$F = B_o \int dz\, \pi z r^2(z) + \int dz\, 2\pi r \sqrt{1 + r_{,z}^2} - \left(\cos\theta_1^{\text{eff}} A_{S_1} + \cos\theta_2^{\text{eff}} A_{S_2}\right) + U_{s0}^{\text{eff}}$$

or

$$F = \int dz \left[B_o \pi z r^2(z) + 2\pi r \sqrt{1 + r_{,z}^2} - \cos\theta_1^{\text{eff}} \pi r^2(z) \delta(z) - \cos\theta_2^{\text{eff}} \pi r^2(z) \delta(z - h_0)\right] + U_{s0}^{\text{eff}},$$

where in $r(z)$ the dependence with $z$ has been omitted for simplicity, and where $B_o = \Delta\rho g l_0^2 / \gamma_{LA}$ is the Bond number. In order to minimize $F$ with constrained droplet volume, we introduce a (dimensionless) Lagrange multiplier $\lambda$

$$F_V = F - \lambda(V - 1), \qquad (S13)$$

where $V$ is the dimensionless droplet volume, $V$

$$V = \pi \int dz\, r^2. \qquad (S14)$$

In order to determine the equilibrium configurations related to the functional Eq. (S12), we derive the Euler-Lagrange equation from (S13) leading to the droplet shape differential equation

$$\sqrt{1 + r_{,z}^2} - \frac{\partial}{\partial z}\left(r \frac{r_{,z}}{\sqrt{1 + r_{,z}^2}}\right) + B_o z r - \lambda r = 0 \qquad (S15)$$

and boundary conditions (BCs)

$$\frac{r_{,z}(h_0)}{\sqrt{1 + r_{,z}^2(h_0)}} - \cos\theta_2^{\text{eff}} = 0 \qquad (S16)$$

$$\frac{r_{,z}(0)}{\sqrt{1 + r_{,z}^2(0)}} + \cos\theta_1^{\text{eff}} = 0. \qquad (S17)$$

The droplet equilibrium configuration is given by solving Eqs. (S15) (with $V = 1$) and (S16-17) with parameters $r(z)$, $\lambda$, $h_0$, $B_o$, $\theta_1^{\text{eff}}$ and $\theta_2^{\text{eff}}$. Thus, once provided $B_o$, $\theta_1^{\text{eff}}$ and $\theta_2^{\text{eff}}$, $r(z)$ and $\lambda$ can be determined as a function of the separation $h_0$.

On the other side, when the droplet is upper-plate-free, we consider the Eqs. (S15) ($V = 1$) and (S17) (with $r(h_0) = 0$). Note that $r(h_0) \to 0$ implies that $\left(\cos\theta_2^{\text{eff}}\right)^2 \to 1$ and similar for the other case. Again, once provided $B_o$, $\theta_1^{\text{eff}}$, $r(z)$ and $\lambda$ can be determined as a function of the droplet height $h_0$.

It is now interesting to determine the force(s) acting on the plate(s). In particular, we have that the droplet mechanical equilibrium requires that

$$p(0)r(0)^2 - p(h_0)r(h_0)^2 = \frac{\Delta\rho V g}{\pi} + 2\gamma_{\text{LV}}(r(0)\sin\theta_1 - r(h_0)\sin\theta_2),$$

whereas by integrating the fluid momentum equation along the $z$-axis

$$p(0) - p(h_0) = \Delta\rho g h_0,$$

leading to

$$p(0)[r(0)^2 - r(h_0)^2] = -r(h_0)^2 \Delta\rho g h_0 + \frac{\Delta\rho V g}{\pi} + 2\gamma_{\text{LV}}(r(0)\sin\theta_1 - r(h_0)\sin\theta_2)$$

$$p(h_0)[r(0)^2 - r(h_0)^2] = -r(0)^2 \Delta\rho g h_0 + \frac{\Delta\rho V g}{\pi} + 2\gamma_{\text{LV}}(r(0)\sin\theta_1 - r(h_0)\sin\theta_2).$$

Therefore, the total force acting on the bottom plate (assumed positive if repulsive) is

$$F_N(0)/\pi = p(0)r(0)^2 - \gamma_{\text{LV}} 2r(0)\sin\theta_1$$

$$= \frac{r(0)^2}{r(0)^2 - r(h_0)^2}\left[-r(h_0)^2\Delta\rho g h_0 + \frac{\Delta\rho V g}{\pi} + 2\gamma_{\text{LV}}(r(0)\sin\theta_1 - r(h_0)\sin\theta_2)\right] - \gamma_{\text{LV}} 2r(0)\sin\theta_1$$

and

$$F_N(h_0)/\pi = p(h_0)r(h_0)^2 - \gamma_{\text{LV}} 2r(h_0)\sin\theta_2$$

$$= \frac{r(h_0)^2}{r(0)^2 - r(h_0)^2}\left[-r(0)^2\Delta\rho g h_0 + \frac{\Delta\rho V g}{\pi} + 2\gamma_{\text{LV}}(r(0)\sin\theta_1 - r(h_0)\sin\theta_2)\right] - \gamma_{\text{LV}} 2r(h_0)\sin\theta_2.$$

In dimensionless units ($\gamma_{\text{LA}} l_0$) we have

$$F_N(0) = B_o \frac{r_0^2(1 - \pi r_{h_0}^2 h_0)}{r_0^2 - r_{h_0}^2} + 2\pi r_{h_0} r_0 \frac{r_{h_0}\sin\theta_1 - r_0\sin\theta_2}{r_0^2 - r_{h_0}^2} \tag{S18}$$

and

$$F_N(h_0) = B_o \frac{r_{h_0}^2(1 - \pi r_0^2 h_0)}{r_0^2 - r_{h_0}^2} + 2\pi r_{h_0} r_0 \frac{r_{h_0}\sin\theta_1 - r_0\sin\theta_2}{r_0^2 - r_{h_0}^2}.$$

Clearly, $F_N(0) - F_N(h_0) = B_o$ which is the droplet-in-air buoyancy in dimensionless units. We note that when the top plate is missing, Eq. (S18) reduces to $F_N(0) = B_o$ as expected.

The upper plate is the hierarchical AAO modified by PFOTS ((1H,1H,2H,2H-Tridecafluorooctyl)trichlorosilane). The contact angle is approximately 150°, see Fig.SS4. In Fig.SS7 we report the force-displacement predictions for the case where the droplet is squeezed between the microtextured surface (bottom plate) and the superhydrophobic surface (top plate, see Fig.SS5). In particular, in Fig.SS7A we report the force (black line) and relative contact radius (blue line), predicted on the bottom plate, as a function of the top plate relative displacement. The contact radius $R$ is made dimensionless with respect to the sessile contact radius $R_0$ the droplet would have if just deposited on the micro-textured surface.

The adopted relationship between the apparent contact angle on the micro-structured surface and the pressure, which models the CB to Wenzel transition on the micro-textured surface (see Supplementary Section S3) is reported in Fig.SS3B.

## S5. Pull-off experiments with droplet

The adhesive force between water droplets and the surfaces of interest was measured by a high-sensitivity microelectromechanical balance system (Dataphysics DCAT11, Germany). 15 µL water droplets were attached to a hydrophobic metal ring, which was set to approach, contact, and leave the surfaces at a constant speed (Ref. 1). A CCD camera was used to monitor the approach and pull-off of the droplet; examples of the recorded images showing the evolution of the interaction between a droplet and the surfaces under investigation is reported in Fig.SS8A. As shown in Fig.SS8B, once the water droplet is in contact with the surface, the force gradually increases and reaches the maximum before the droplet separates from the surface; the maximum values of the forces obtained for all surfaces were recorded in the curve as the adhesive forces. The residual mass of the water left on the surface is also recorded as reported in Fig.SS8C.

## S6. Tribocurrent generation

Literature reports values of PP surface charging upon distilled water contact of $\rho_A \approx 6 * 10^{-6}$ C/m$^2$. (Refs. 2, 3) Furthermore, in our triboelectric generator, the nominal tank flat area is $A_0 = 1.26 \cdot 10^{-3}$ m$^2$.

In the case of untextured surface, the true water/PP contact area $A_c$ corresponds to $A_0$, leading to an ideal tribo-charge of order $Q_{unt} = A_0 \rho_A = 0.76 \, nC$. The latter is reported in Fig.2(D.1) as a reference value for ideal flat surface. Nevertheless, the drag-out dewetting (see Supplementary Section S9) is not completed at sloshing frequency of 3Hz (i.e., the tank flat surface is not fully freed from a water film upon tank motion reversal), thus leading to a much-reduced tribocurrent generation, see Fig.2(D.1).

For the nanotextured surface the drag-out dewetting occurs, see Fig.2(D.2). In this case $A_c = A_0 \alpha_c^N \approx 0.57 A_0$, leading to a tribocurrent $Q_N \approx 0.43$ nC, in agreement with the experimental data, see Fig.2(D.1).

For the hierarchical texture, $A_c = A_0 \alpha_c^H$, where $\alpha_c^H = \alpha_c^N * \alpha_c^m * a$, where $\alpha_c^m \approx 0.33$ and $a \approx 6.3$ (considering the largest microtexture wavelength to be sinusoidal, with same wavelength and amplitude, see e.g. Fig.1(B.1)). This leads to $Q_H \approx 0.89$ nC, in agreement with the experimental results.

## S7. TENG experiments at different sloshing frequencies

Fig.SS9A shows a sequence of fluid motion images of the water-based triboelectrification power generation cycle for typical experimental parameters adopted during the testing: water-to-cylinder volume ratio 0.3, vibration frequency 3 Hz, and vibration amplitude 5 mm. The outputs generated in terms of current and voltage are shown in Figs.SS9B and SS9C respectively.

As already shown in our previous preliminary contributions (Ref. 4) describing the response of an optimized U-shaped triboelectrification device, the output current and voltage can show some degree of asymmetry. This might depend on the local conditions and the accuracy of the system design in terms of vibrations, as the system is particularly sensitive to the sloshing dynamics as discussed in the main manuscript. However, this aspect is not the focus of this contribution, whose main aim is to describe the remarkable ability of triboelectrification to finely detect the wetting dynamics.

## S8. Weak non-linear theory of sloshing dynamics

In this section we report the model describing the sloshing dynamics characterizing the water-filled cylindrical-tank triboelectric generator. The contact pressure acting on the functionalized nominally-flat cylinder surfaces can be approximately predicted recurring to a weakly-nonlinear sloshing dynamics model of the partially filled tank. We note that the liquid free-surface (in partially filled tanks) can experience a wide cascade of energetic phenomena and related kinematics, the latter ranging from the simplest planar motion, to no-planar, rotational, quasi-periodic, random and lasting with free-surface fragmentation processes. A comprehensive review on the sloshing dynamics motions can be extracted by Ref. (5).

Here we focus on the determination of the average water squeezing pressure acting on the cylinder flat surfaces. To do so, we model the sloshing vibration dynamics within a Rayleigh approach by assuming the free-surface motion to be governed by a simple two-dimensional kinematics, where the free surface is approximated to a flat tilting surface, with rotation axis perpendicular to the cylinder tank axis of symmetry, see Fig.SS10(A.1). The application of the mass conservation (volume conservation in this case) leads the free-surface kinematics to be described by just one free Lagrangian parameter, which we consider to be the tilting angle $\varphi$ of the free-surface with respect to the cylinder axis (with $\alpha = tan(\varphi)$).

In the following we report the main equations describing the surface kinematics. In particular, by assuming lengths (such as the cylinder tank axial length, L$_0$) **dimensionless** with respect to the cylinder radius, $R_c$, the relation between the center of mass $\mathbf{x}_G = (x_G, z_G)$ as a function of the tilting angle $\alpha = tan(\varphi)$ is calculated as follows:

$$z_G = \frac{-\frac{2\sin^{-1}(1+\alpha l_1)+\pi}{8\alpha} + \frac{(2\alpha^3 l_1^3 + 6\alpha^2 l_1^2 + \alpha l_1 - 3)\sqrt{\alpha(-l_1)(\alpha l_1 + 2)}}{12\alpha}}{V}$$

$$\text{if } 2 + l_1\alpha < L_0\alpha \text{ and } l_1 < 0$$

$$z_G = \frac{-\frac{\pi}{4\alpha}}{V}$$

$$\text{if } 2 + l_1\alpha < L_0\alpha \text{ and } l_1 \geq 0$$

$$z_G = -\frac{(1+\alpha(l_1-L_0))(2(1+\alpha(l_1-L_0))^2 - 5)\sqrt{1-(1+\alpha(l_1-L_0))^2}}{12\alpha V} +$$
$$+ \frac{\sin^{-1}(1+\alpha(l_1-L_0)) - \sin^{-1}(1+\alpha l_1)}{4\alpha V} + \frac{(\alpha l_1 + 1)(2\alpha^2 l_1^2 + 4\alpha l_1 - 3)\sqrt{\alpha(-l_1)(\alpha l_1 + 2)}}{12\alpha V}$$

$$\text{if } 2 + l_1\alpha \geq L_0\alpha \text{ and } l_1 < 0$$

$$z_G = -\frac{(1+\alpha(l_1-L_0))(2(1+\alpha(l_1-L_0))^2 - 5)\sqrt{1-(1+\alpha(l_1-L_0))^2}}{12\alpha V} + \frac{\sin^{-1}(1+\alpha(l_1-L_0)) - \frac{\pi}{2}}{4\alpha V}$$

$$\text{if } 2 + l_1\alpha \geq L_0\alpha \text{ and } l_1 \geq 0,$$

and

$$x_G = \frac{(\alpha l_1 + 1)(1+\alpha(l_1-L_0))\tan^{-1}\left(\frac{\alpha(-l_1)-1}{\sqrt{1-(\alpha(-l_1)-1)^2}}\right) - \pi(2+\alpha l_1)(2+\alpha(l_1-L_0))}{2\alpha^2 V} +$$
$$+ \frac{6\pi(\alpha l_1 + 1)(1+\alpha(l_1-L_0)) + 12\pi(\alpha(-L_0) + 2\alpha l_1 + 3)}{24\alpha^2 V}$$
$$+ \frac{-3\sin^{-1}(\alpha l_1 + 1) - \frac{3\pi}{2} - b_1\sqrt{\alpha(-l_1)(\alpha l_1 + 2)}}{24\alpha^2 V}$$

$$\text{if } 2 + l_1\alpha < L_0\alpha \text{ and } l_1 < 0$$

$$x_G = \frac{\left\{\frac{\pi(-4\alpha L_0 + 8\alpha l_1 + 11)}{8\alpha^2} - \frac{\pi(2+\alpha l_1)(2+\alpha(l_1-L_0))}{2\alpha^2}\right\}}{V}$$

$$\text{if } 2 + l_1\alpha < L_0\alpha \text{ and } l_1 \geq 0$$

$$x_G = \frac{12(\alpha l_1 + 1)(1+\alpha(l_1-L_0))\left(\tan^{-1}\left(\frac{\alpha(-l_1)-1}{\sqrt{1-(\alpha(-l_1)-1)^2}}\right) - \tan^{-1}\left(\frac{(1-\alpha(l_1-L_0))}{\sqrt{1-(1-\alpha(l_1-L_0))^2}}\right)\right)}{24\alpha^2 V}$$
$$+ \frac{a_1\sqrt{1-(1+\alpha(l_1-L_0))^2} - b_1\sqrt{\alpha(-l_1)(\alpha l_1 + 2)} + 3\sin^{-1}(\alpha(-L_0) + \alpha l_1 + 1) - 3\sin^{-1}(\alpha l_1 + 1)}{24\alpha^2 V}$$

$$\text{if } 2 + l_1\alpha \geq L_0\alpha \text{ and } l_1 < 0$$

$$x_G = \left\{\frac{12(\alpha l_1 + 1)(\alpha L_0 - \alpha l_1 - 1)\tan^{-1}\left(\frac{(1-\alpha(l_1-L_0))}{\sqrt{1-(1-\alpha(l_1-L_0))^2}}\right) + a_1\sqrt{1-(1+\alpha(l_1-L_0))^2}}{24\alpha^2 V}\right.$$

$$\left. + \frac{-6\pi(\alpha l_1 + 1)(1+\alpha(l_1-L_0)) + 3\sin^{-1}(1+\alpha(l_1-L_0)) - \frac{3\pi}{2}}{24\alpha^2 V}\right\}$$

$$\text{if } 2 + l_1\alpha \geq L_0\alpha \text{ and } l_1 \geq 0,$$

where $V$ is considered fixed when assuming the liquid to be incompressible, with the volume fraction of cylinder filled with water $V_{\text{frac}} = V/(\pi L_0)$. Furthermore, we have that

$$a_1 = \alpha(2\alpha^2(L_0 - l_1)^2(L_0 + l_1) - 2\alpha(L_0 - l_1)(L_0 + 3l_1) - 7L_0 + 19l_1) + 15$$
$$b_1 = (-4\alpha L_0(\alpha^2 l_1^2 + 2\alpha l_1 + 3) + 2\alpha^3 l_1^3 + 6\alpha^2 l_1^2 + 19\alpha l_1 + 15).$$

and $l_1$ is given by:

$$V = \frac{\pi}{\alpha} + \frac{(1+\alpha l_1)\cos^{-1}(1+\alpha l_1)}{\alpha} - \frac{(3+\alpha l_1(2+\alpha l_1))\sqrt{\alpha(-l_1)(2+\alpha l_1)}}{3\alpha} + \pi\left(L_0 - \left(\frac{2}{\alpha}+l_1\right)\right)$$

if $2 + l_1\alpha < L_0\alpha$ and $l_1 < 0$,

$$V = \frac{\pi}{\alpha} + \pi\left(L_0 - \left(\frac{2}{\alpha}+l_1\right)\right)$$

if $2 + l_1\alpha < L_0\alpha$ and $l_1 \geq 0$,

$$V = \frac{[2+(1+\alpha(l_1-L_0))^2]\sqrt{1-(1+\alpha(l_1-L_0))^2}}{3\alpha} - \frac{(2+(1+\alpha l_1)^2)\sqrt{\alpha(-l_1)(2+\alpha l_1)}}{3\alpha}$$
$$+ \frac{(1+\alpha l_1)\cos^{-1}(1+\alpha l_1) + (1-\alpha(l_1-L_0))\cos^{-1}(1+\alpha(l_1-L_0))}{3\alpha}$$

if $2 + l_1\alpha \geq L_0\alpha$ and $l_1 < 0$,

$$V = \frac{\left(3-\alpha(L_0-l_1)(2+\alpha(l_1-L_0))\right)\sqrt{\alpha(L_0-l_1)(2+\alpha(l_1-L_0))}}{3\alpha}$$
$$- \frac{(1+\alpha(l_1-L_0))\cos^{-1}(1+\alpha(l_1-L_0))}{\alpha}$$

if $2 + l_1\alpha \geq L_0\alpha$ and $l_1 \geq 0$.

We observe that, by varying the tilting angle $\varphi$, the fluid center of mass $\mathbf{x}_G$ decribes a trajectory $\mathbf{x}_G(\alpha)$, whose curvilinear abscissa we name $s(\alpha)$. In the model presented here, mass conservation implies that the trajectory tangent $d\mathbf{x}_G/ds$ is given with $dz_G(s)/dx_G(s) = \tan\alpha$ (it can be easily proved by evaluating the variation of $(\mathbf{x}_G V)_{|s}$). Finally, assuming all the liquid inertia concentrated in the center of mass, the linear momentum conservation imposes that

$$\ddot{s}(t) + c\dot{s}(t) + g\sin\alpha(t) + \ddot{x}_0(t)\cos\alpha(t) = 0, \tag{S19}$$

where $\alpha(t) = \alpha(s(t))$, and where again lenghts are made dimensionless using the cylinder radius as reference length, and time using $(2\pi f)^{-1}$, where $f$ is the excitation frequency of the external shaking motion of the tank which reads (in dimensionless units) $x_0(t) = x_0\sin(t)$. $g$ is the dimensionless gravitational acceleration $[gR^{-1}(2\pi f)^{-2}]$, assumed perpendicular to the cylinder axis. $c$ is a (dimensionless) fluid damping parameter which can be evaluated as $c \approx \nu/(2\pi f h_0^2)$, where $\nu$ is the fluid kinematic viscosity, and where $h_0$ is representative of the sheared fluid thickness. In our system, $2\pi f c \approx 10^{-2}$ to $10^{-1}\text{s}^{-1}$ for a tank filling ratio of 30% (value adopted in the experiments).

Eq. (S19) is evaluated numerically, upon which the (dimensionless) fluid squeezing load acting on the tank flat surfaces can be calculated, within the approximated model above, as follows

$$F_w(t) = \left[g\cos\alpha(t) + \frac{\dot{s}(t)^2}{r(s(t))}\right]\sin\alpha(t),$$

where $r(s)$ is the curvature radius of the trajectory at abscissa $s$

$$r(s)^{-1} = \frac{d\tan\alpha(s)}{ds}(\cos\alpha(s))^2.$$

Finally, the fluid squeezing pressure, with dimensions, for our system is $p_w = \frac{1}{2}c_w\rho\bar{v}^2$, where $\bar{v} = 2fL_0$ is the average sloshing speed and $c_w = 2\pi^2 F_w R L_0^{-1} V_{\text{frac}}$ is the dimensionless wall impact coefficient ($V_{\text{frac}}$ is the tank filling ratio). Results for the case of $f = 3\,Hz$ are reported in Fig.2(B.2), whereas the sloshing trajectories at varying filling ratios are reported in Fig.SS10(A.2).

### S9. Film dewetting by drag-out mechanics

The sloshing motion causes the bulk waterfront to be squeezed/retrieved onto/from the tank flat surfaces, nominally at the same frequency of the shaking motion. However, after this bulk detachment of the waterfront, a thin water film might be left on the flat surfaces. Indeed, the dewetting time, necessary for this residual fluid film to be drag-out from the flat surfaces, is not necessarily matching the sloshing frequency, and, as such, it needs to be evaluated by separately studying the drag-out dewetting dynamics of the water front on the flat cylinder surfaces. Interestingly, if the characteristics time of the drag-out dynamics is larger than the characteristics sloshing time, then the water film is never dewetting from the PP surfaces, leading to no triboelectric charging in the system.

Here we evaluate the drop off thickness of the water on the PP plate assuming the confined liquid film removal to be described within the so-called drag-out problem, see Fig.SS11, where an inclined plate is withdrawn from a pool of liquid and one needs to calculate the thickness of the film clinging to the plate. For a vertical plate, this setting was examined by Landau and Levich (1942), a theory later extended by Wilson (1982) to the case of a plate inclined at an arbitrary angle (so called LLW solution). Here we use the non-LLW solutions to describe the general case of non-perfectly wetting liquid, where the film thickness is predicted to be $h_{\min} = 3^{1/2}\sqrt{v_0 \eta/(\rho g)}$, with $v_0 = 2Rf$ the drag-out speed. Thus, for our system, we get $h_{\min} = 0.110, 0.156$, and $0.192$mm at the different sloshing frequencies of the experiments, respectively, 1, 2 and 3 Hz.

Once knowing the film thickness $h_{\min}$ during drag-out, we now need to determine whether this water layer is energetically stable or either a dewetting transition is more favorable to occur, as well its characteristics dewetting time. To do so, we make use of the thin film (lubrication) theory. In particular, we calculate first the disjoining pressure $\Pi(h)$ acting in the film (which can be obtained from the Gibbs free energy $G(h)$ of the film, with $\Pi(h) = -dG(h)/dh$) whose polar and apolar contributions are

$$\Pi(h) = 2S_{\text{AP}}\frac{d_0^2}{h^3} + \frac{S_{\text{P}}}{l}e^{\frac{d_0-h}{l}}, \tag{S20}$$

where $d_0 = 0.158$ nm is the Born repulsion length and $l$ the correlation length for a polar fluid, which for water is about $l_W = 0.6$ nm. $S_P$ and $S_{AP}$ are the polar and apolar components of the spreading coefficient $S = S_P + S_{AP}$. We also have that $S = \gamma_{SV} - \gamma_{SL} - \gamma_{LV} = \gamma_{LV}(\cos\theta - 1)$, where the thermodynamic contact angle of water on pristine PP $\theta = 102°$. The apolar component $S_{AP}$ can be derived from the effective Hamaker constant of the interface air-water-solid (PP is the solid), $A_{\text{GWS}}$, with $S_{\text{AP}} = -A_{\text{GWS}}/(12\pi d_0^2)$. For contact of two dissimilar materials in the presence of a third media $A_{\text{GWS}} = A_{\text{GS}} + A_{\text{WW}} - A_{\text{GW}} - A_{\text{WS}} \approx A_{\text{WW}} - A_{\text{WS}}$. Since $A_{\text{WS}} \approx \sqrt{A_{\text{WW}}A_{\text{SS}}}$ one has

$$A_{\text{GWS}} \approx \sqrt{A_{\text{WW}}}(\sqrt{A_{\text{WW}}} - \sqrt{A_{\text{SS}}}).$$

We note further that $A_{\text{SS}} = 24\pi d_0^2 \gamma_S$. For polypropilene, $\gamma_S = 0.031$N/m, whereas for water $A_{\text{WW}} = 4.38 \times 10^{-20}$Nm, leading to $A_{\text{SS}} = 5.83 \times 10^{-20}$Nm and to $A_{\text{GWS}} = -6.75 \times 10^{-21}$Nm. Thus, we get $S_{\text{AP}} = 0.0072$N/m and $S_{\text{P}} = -0.094$N/m (Ref. 6). Therefore, in the disjoining pressure the apolar component is stabilizing ($S_{\text{AP}} > 0$) whereas the polar is destabilizing ($S_{\text{P}} < 0$) the water film thickness. In Eq. (S20), for $h$ larger than $\approx 0.265$ nm we have $d\Pi(h)/dh > 0$, i.e. the water films are nominally unstable in our case.

If we assume the film breakdown occurs in correspondence of the discontinuities of the tank flat surfaces, we can now easily estimate the time needed for the film to retrieve from the flat surfaces. Within lubrication hydrodynamics this characteristic time $t_0$ reads

$$t_0 \approx \frac{6s_0^2 \eta}{h_{\min}^2 (p_0 - 2h_{\min}^{-1}\gamma_{\text{LV}}\cos\theta_{eff})},$$

with $p_0 = gs_0\rho$ and $s_0 = 2R$, and where $\theta_{eff}$ is the apparent water contact angle on the surface.

It is interesting to compare the case of dewetting occurring on the pristine PP surface (untextured surface) with the nanotextured surface. Indeed, since the water-PP contact area is reduced in the nanotextured case with respect to the untextured case, one would expect the surface charging to be higher for the untextured surface. Nevertheless, we get that for the case of $\theta_{eff} = 102°$ (pristine PP surface) the water film dewets the PP surface only for frequencies lower than $\approx 0.5$Hz, whereas for $\theta_{eff} = 123°$ (nanotextured surface) for frequencies lower than $\approx 3.4$Hz, see Fig.2(D.2). Thus, for the untextured case, the water is not able to fully dewet the PP surface in the range of sloshing frequencies under investigation, remarkably leading to a smaller charge accumulation with respect to the nanotextured case.

## S10. TENG experiments at different surface energies

In a liquid, the variation of the surface tension can also drastically influence the Cassie-Baxter-to-Wenzel transition, thus the triboelectrification. This can be proved by simply adding surfactants to change the surface tensions of water. Therefore, after analyzing the effect of different surfaces and sloshing frequencies as reported in Fig.2 and Fig.SS12, the effect of surfactant contamination was investigated. Fig.SS13 shows that the output current of water-based triboelectrification device is greatly related to the surfactant concentration. It could be observed that the current stays at the maximum level (~40 nA) when there is no surfactant concentration, and decreases to a lower level (~0.4 nA) once the surfactant is added to the system. Fig.SS13C presents the sensitivity of a water-based hierarchical triboelectrification device and a water-based smooth triboelectrification device in the detection of surfactants. It is shown that the output current of water-based hierarchical triboelectrification device decreases more drastically as the surfactant concentration is increased. Compared to a water-based smooth triboelectrification device, water-based hierarchical triboelectrification devices have a better sensitivity to detect the presence of surfactants as their addition can change the wetting state and the concentration of ions significantly. The decrease of the surface triboelectric charges of the water-based smooth triboelectrification device only contributes to an increase of the concentration of ions. In contrast to this, for the water-based hierarchical triboelectrification device the wetting state transition of the polymeric surface also affects triboelectric generation. So, the water-based hierarchical triboelectrification device shows superiority in the detection of surfactant.

With the surfactant concentration changing from 0 to 1μM, the generated current of the water-based hierarchical surface does not register particular reductions. When the surfactant concentration changes from 10 μM to 100 μM, the generated current marginally decreases. However, the generated current shows a faster decay when the surfactant concentration increases from 500 μM to 1 mM, see Fig.SS13C. Thus, the relationship between the output of water-based triboelectrification device and surfactant concentration could be divided into three wetting states (CB state, partially Wenzel state, and full Wenzel state) based on the magnitude of the current reduction as shown in Fig.SS13. In the low surfactant concentration (CB state), the polymer surface maintains superhydrophobicity during the wetting/dewetting cycles. The output current of the water-based hierarchical triboelectrification device has little change as time increase. The decrease of output current with the increase of surfactant concentration merely contributes to the increase of ions concentration. In the partially wetted state, part of the PP surface has been wetted, and the wetting state starts to transform the response from the Cassie-Baxter state to the Wenzel state. Since a permanent water film is formed, the output current of the water-based hierarchical triboelectrification device marginally decreases, then remains unchanged. In the full Wenzel state, the polymer surface is irreversibly wetted. The output current shows a faster decay during the contact-separation (wetting-dewetting) cycles. The Cassie-Baxter-to-Wenzel transition and the increase of ions concentration act synergistically to cause the change of current output.

## **S11. Theory of multiscale wetting film formation on rough surface**

In this section we report a novel theory, based on the random walk process, for the prediction of the statistics of wetting film thickness on a generic random multiscale surface roughness. We observe that, upon deposition of a thin liquid film (with initial uniform thickness $h$) on a smooth rigid substrate, spinoidal dewetting occurs for $d\Pi(h)/dh > 0$, where $\Pi(h)$ is the disjoining pressure (see e.g. Supplementary Section S9). For our system (polypropilene), instability occurs for thin films with thickness larger than $h_{\lim} = 0.26 \text{ nm}$ on an ideally smooth substrate.

We now consider the case of a thin uniform film of initial thickness $v_0$ on a rough rigid surface, characterized by a single wavelength $q$ and amplitude $h = h(q)$, *i.e.* with surface roughness $h \sin(qx)$. In such a case, dewetting occurs accordingly to a thin film evolution equation for the amplitude (linearized model, see e.g. Ref. 7) which reads

$$v(t) = v_0 - hq^2(q^2 - k_c^2)^{-1}\left(1 - e^{-\frac{t}{\tau}}\right), \tag{S21}$$

where $v(t)$ is the film thickness at time $t$, and $\tau$ is a time scale for dewetting ($\tau \propto (q^2 - k_c^2)^{-1}$). $k_c^2 = \Pi'(v_0)/\gamma_l$ is the critical spinoidal wavelength, obtained by linearizing the model around the initial film thickness value $v_0$. In the simplest picture, roughness-driven dewetting and spinoidal dewetting are the two competing mechanisms delivering the breakdown of an energetically unstable film ($d\Pi(h)/dh > 0$). We note however that the fastest dewetting mode is also strongly affected by the initial inhomogeneity of the film thickness [7]. To simplify the model, in the following we assume the dewetting to occur independently at each roughness wavelength, consistently with the Gaussian nature of the surface roughness at lengthscales larger than the mean microcube size, see Fig.SS14.

We now consider the surface to be characterized by a random roughness $h(\mathbf{q})$, where $h(\mathbf{q}) = (2\pi)^{-2} \int d^2x\, h(\mathbf{x}) e^{-i\mathbf{q}\cdot\mathbf{x}}$ and $h(\mathbf{x})$ is the random surface roughness. We observe the surface at magnification $\zeta = q/q_0$ ($q = |\mathbf{q}|$), i.e. by ideally including roughness up to the frequency $q$ (where $q_0 = 2\pi/L_0$ is a reference frequency, say the roughness lower cutoff frequency of the power spectral density, PSD, see Fig.3(A.1)). At this magnification $\zeta$ a fraction of the total surface $A_0$ surface (say $A_w(\zeta)$) appears covered by a random distribution of film thicknesses with average volume $V_0 = v_0 A_0$, as a consequence of the dewetting due to the higher roughness wavelengths $\lambda > 2\pi/q$. $A_w(\zeta)$ is the apparent wet surface area at magnification $\zeta$, i.e. the wet area obtained when adding roughness up to the frequency $q = \zeta q_0$. At magnification $\zeta$, the probability distribution of wetting film thickness is $p(q = \zeta q_0, v)$, with $A_w(\zeta) = \int_0^\infty dv\, p(\zeta, v)$ and with $v(\mathbf{x}, q)$ the film thickness distribution. We now add a small amount of (random) surface roughness in the frequency range $[q, q + \Delta q]$, where $\Delta q = q_0 \Delta \zeta$ is the surface roughness frequency variation. Therefore, when we add this small amount of smaller scale roughness with wavenumber between $q$ to $q + \Delta q$, $v(\mathbf{x}, q)$ randomly grows by an amount $\Delta v(\mathbf{x}, q)$. We assume $\Delta v(\mathbf{x}, q)$ to be the sum of an infinite number of wavevectors $\mathbf{q}$ in the range $q$ to $q + \Delta q$ with homogeneously distributed phases, thus statistically independent wave-components. Therefore, $\Delta v(\mathbf{x}, q)$ is a Gaussian variable, with probability density function $\text{p}_{\Delta v(\mathbf{x},q)}$ given by

$$\text{p}_{\Delta v(\mathbf{x},q)}(\Delta v) = \frac{e^{-\frac{\Delta v^2}{2\langle\Delta v(\mathbf{x},q)^2\rangle}}}{\sqrt{\pi}\sqrt{2\langle\Delta v(\mathbf{x},q)^2\rangle}}. \tag{S22}$$

In order to calculate the probability density function $p(q + \Delta q, v)$ of the film thickness field $v(\mathbf{x}, q + \Delta q)$ when adding roughness in the range $q$ to $q + \Delta q$, we can use Bayesian inference, resulting in

$$p(q + \Delta q, v) = \int_0^\infty dv'\, \text{p}_{\Delta v(\mathbf{x},q)}(v - v') p(q, v'). \tag{S23}$$

Hereinafter we omit the dependence on the spatial variable $x$. After some manipulation, we write (S23) as

$$\text{p}(q + \Delta q, v) = \int_{-\frac{v}{(2\langle\Delta v(\mathbf{x},q)^2\rangle)^{1/2}}}^\infty dz\, \frac{e^{-z^2}}{\sqrt{\pi}} p\big(q, v + z\sqrt{2\langle\Delta v(\mathbf{x}, q)^2\rangle}\big). \tag{S24}$$

We then expand $\text{p}\big(q, v + z\sqrt{2\langle\Delta v(\mathbf{x}, q)^2\rangle}\big)$ in $\sqrt{2\langle\Delta v(\mathbf{x}, q)^2\rangle}$ to get

$$\text{p}\big(q, v + z\sqrt{2\langle\Delta v(\mathbf{x}, q)^2\rangle}\big) = \text{p}(q, v) + \frac{\partial \text{p}(q, v)}{\partial v}\langle 2\Delta v(q)^2\rangle^{1/2} z$$
$$+ \frac{\partial^2 \text{p}(q, v)}{\partial v^2}\frac{\langle 2\Delta v(q)^2\rangle z^2}{2} + o\big(\langle 2\Delta v(q)^2\rangle^{3/2} z^3\big),$$

and $\text{p}(q + \Delta q, v)$ in $\Delta q$ to get

$$\text{p}(q + \Delta q, v) = \text{p}(q, v) + \frac{\partial \text{p}(q, v)}{\partial q}\Delta q + o(\Delta q^2).$$

In the limit $\Delta q \to 0$ (S24) becomes

$$\text{p}(q, v) + \frac{\partial \text{p}(q, v)}{\partial q}\Delta q = \int_{-\infty}^\infty dz\, \frac{e^{-z^2}}{\sqrt{\pi}}\left[\text{p}(q, v) + \frac{\partial \text{p}(q, v)}{\partial v}\langle 2\Delta v(q)^2\rangle^{1/2} z + \frac{\partial^2 \text{p}(q, v)}{\partial v^2}\cdot\frac{\langle 2\Delta v(q)^2\rangle z^2}{2}\right]$$

which, after integration and defining $\langle\Delta v(q)^2\rangle = \Delta q\, 2f(q)$, becomes

$$\frac{\partial p(q,v)}{\partial q} = f(q) \frac{\partial^2 p(q,v)}{\partial v^2}, \tag{S25}$$

where $v(q)$ is the diffusion term. (S25) is solved with the initial condition $p(q_0, v) = \delta(v - v_0)$, where $v_0$ is the initial average film thickness, and with $p(q, v = 0) = p(q, v \to \infty) = 0$. This leads to the solution

$$P(F, v) = \frac{e^{-\frac{(v-v_0)^2}{4F}}}{2\sqrt{\pi F}} - \frac{e^{-\frac{(v+v_0)^2}{4F}}}{2\sqrt{\pi F}}, \tag{S26}$$

where $dF(q)/dq = f(q)$. The normalized projected wet area

$$\frac{A_w}{A_0} = \int dv\, P(F, v) = \text{erf}\left(\frac{v_0}{2\sqrt{F}}\right) \tag{S27}$$

and the volume of fluid conservation is satisfied

$$\int dv\, vP(F,v) = v_0.$$

We now need to calculate $\langle \Delta v(q)^2 \rangle$. From (S21)

$$v(q) = h(q)\frac{q^2}{q^2 - k_c^2}$$

leading to

$$\langle \Delta v(q)^2 \rangle = 2\pi q \left(\frac{q^2}{q^2 - k_c^2}\right)^2 C(q)\, dq = \Delta q\, 2f(q)$$

and to

$$F(q) = \int_0^q d\bar{q}\, \pi\bar{q} \left(\frac{\bar{q}^2}{\bar{q}^2 - k_c^2}\right)^2 C(\bar{q}).$$

If we consider the case where $\bar{q} \ll k_c$, then $F(q) = k_c^{-4}\int_0^q d\bar{q}\, \pi\bar{q}^5 C(\bar{q}) = k_c^{-4} m_4(q)/2$, *i.e.* proportional to the scale-dependent mean square curvature $m_4(q)$. The true normalized projected wet area is therefore

$$\frac{A_w(q_1)}{A_0} = \int dv\, P(F(q_1), v) = \mathrm{erf}\left(\frac{v_0}{2\sqrt{F(q_1)}}\right) \qquad (S28)$$

with a probability distribution of film gap

$$P(q_1, v) = \frac{e^{-\frac{(v-v_0)^2}{4(q_1)}}}{2\sqrt{\pi(q_1)}} - \frac{e^{-\frac{(v+v_0)^2}{4(q_1)}}}{2\sqrt{\pi(q_1)}}, \qquad (S29)$$

where $q_1$ is the roughness high frequency cut-off.

In Fig.SS15 we report the epifluorescence optical acquisitions of the residual wet areas on the textured PP surfaces at different sloshing times (3Hz sloshing frequency). Rodhamine B is dissolved in water with concentration 0.01mg/ml. The acquisitions (lighting and image acquisition) have been triggered in a reduced set of time intervals, thus avoiding any significant photo-bleaching phenomenon to occur in the experiments. Comparisons between the measured residual intensity and the predicted results is terms of probability distribution of film gap are reported in the main text and Fig.3.

**Fig.SS1.**

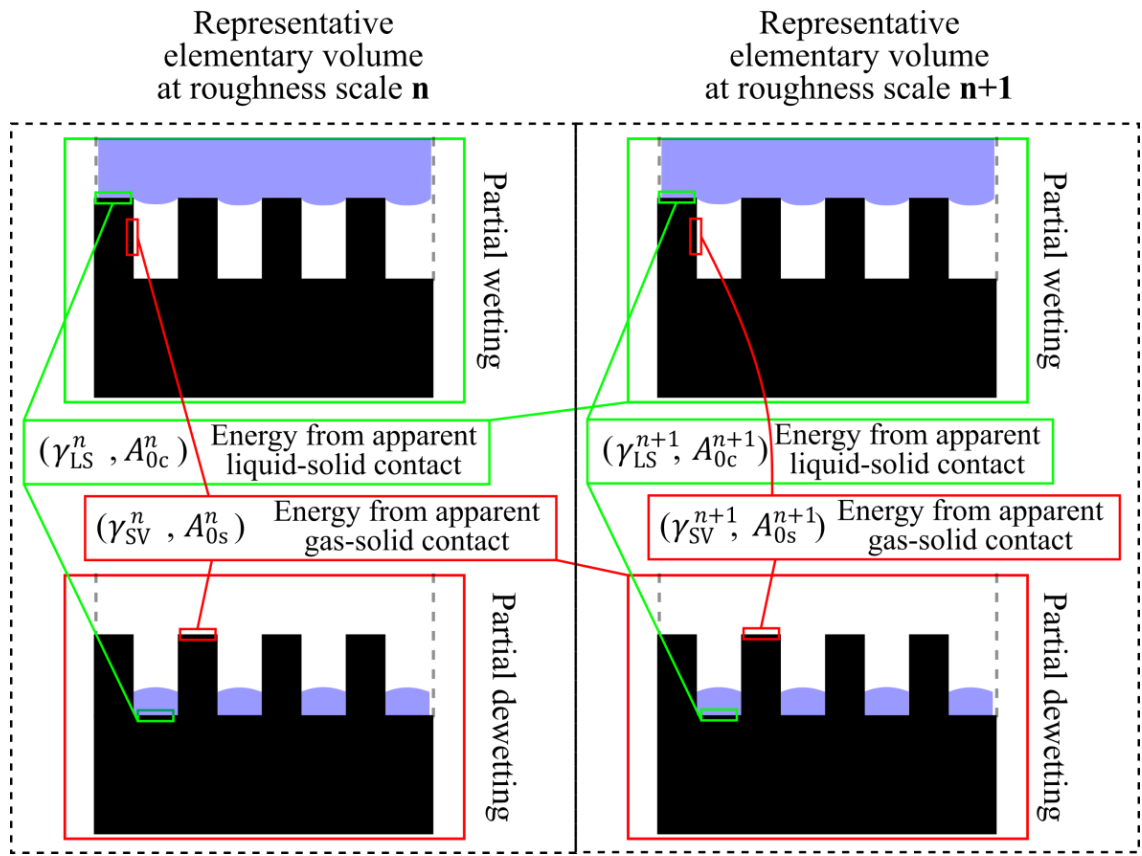

Schematic representation of the generic multiscale wetting process, involving different magnifications in the contact geometry (such as the length scale *n* and *n+1*, in the figure above). At each magnification, the contact can be in a partial wetted (green box) or de-wetted (red box) state.

**Fig.SS2.**

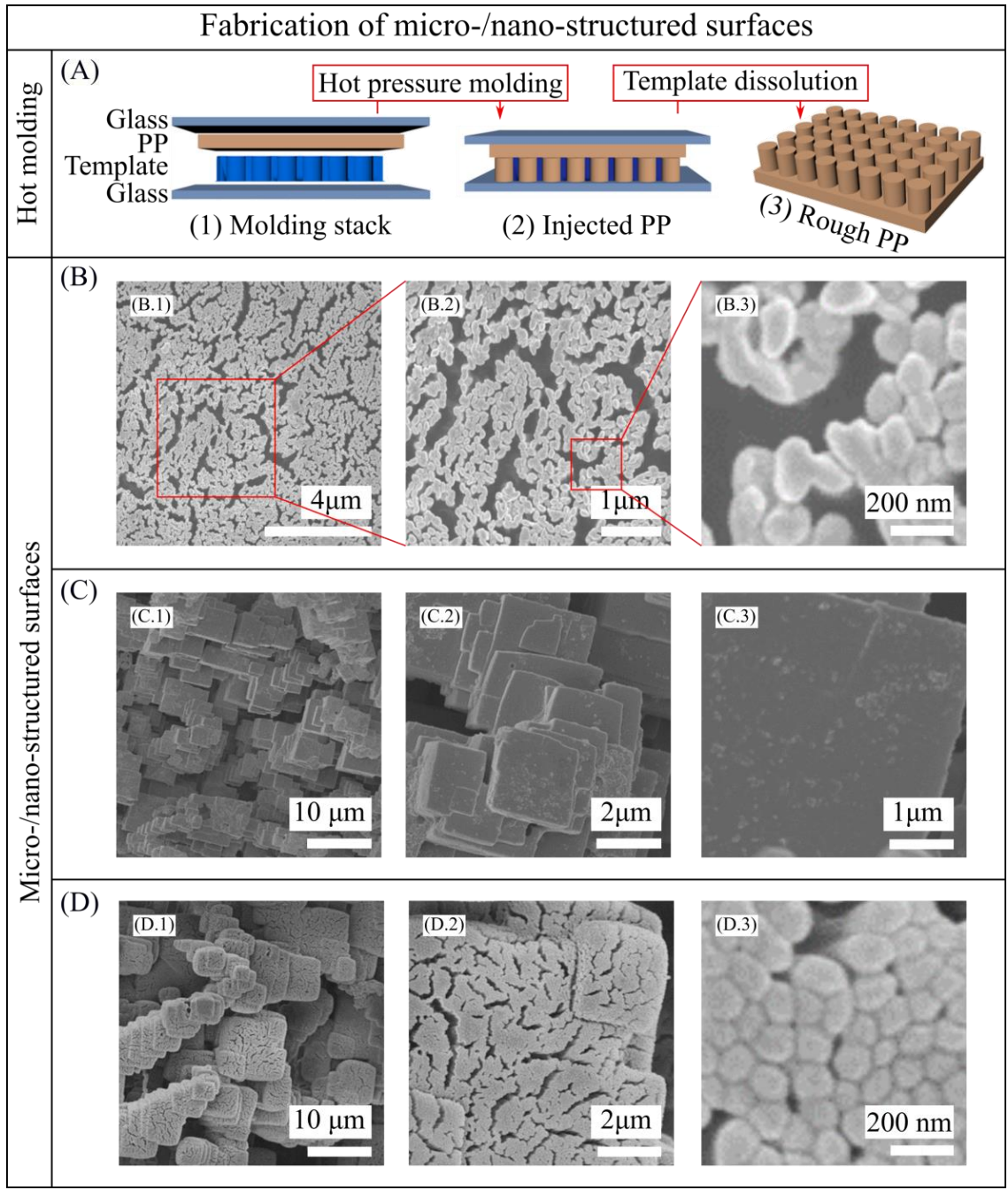

Schematics of the fabrication process of the different surface roughness on PP substrate. (A) a fabrication process of the PP based on a hot molding process. (B) FESEM images of the nanotextured PP, (C) microtextured PP and (D) the hierarchical PP. (D.3) FESEM micrographs of the nanopillar forest covering the micro-cubes.

**Fig.SS3.**

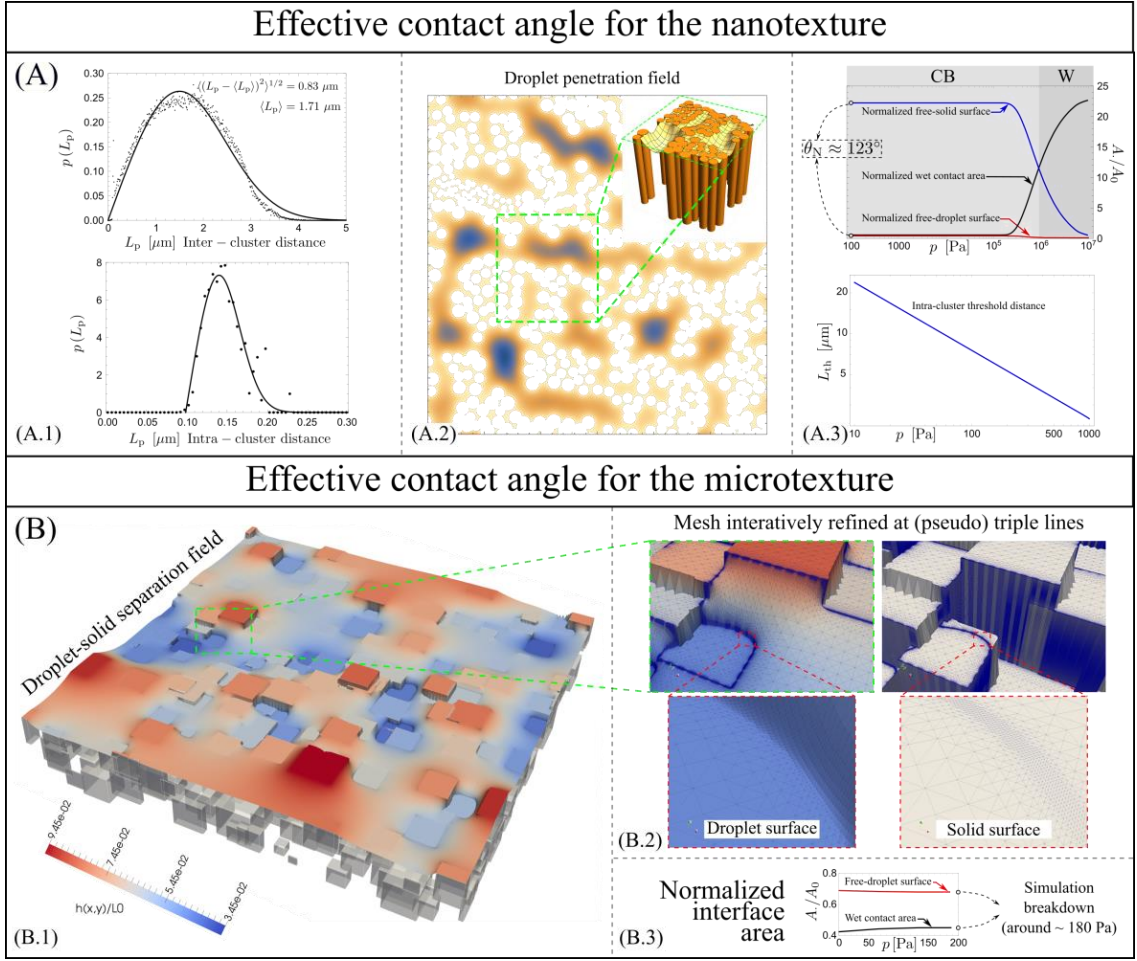

Wet contact simulation results. (A) The case of nanopillar surface, with indication of (A.1) inter- and intra-cluster distance probability density function. (A.2) Example of water penetration into the nanopillar forest with related three-dimensional representation (exaggerated penetration values). (A.3) effective contact angle (top) and intra-cluster threshold distance as a function of the water squeezing pressure. A Cassie-Baxter (CB) regime applies in the range of pressures of relevance for our application (see text), whereas the Wenzel regime is gradually experienced only for pressures larger than order $10^5$ Pa, however the reversible wet contact area is constant, and given by the cumulative top pillars area. The inter-cluster threshold distance, as a function of water squeezing pressure, is the mean lattice spacing between clustered nanopillars at which the Wenzel state occurs. (B) The case of microtextured surface. (B.1) Example of water surface contacting the microtextured surface, with magnified view of the (B.2) water front and microtextured surface meshes. (B.3) Free-water surface and wet contact normalized surface area as a function of the squeezing pressure. The model breaks down upon reaching a certain amount of squeezing pressure.

**Fig.SS4.**

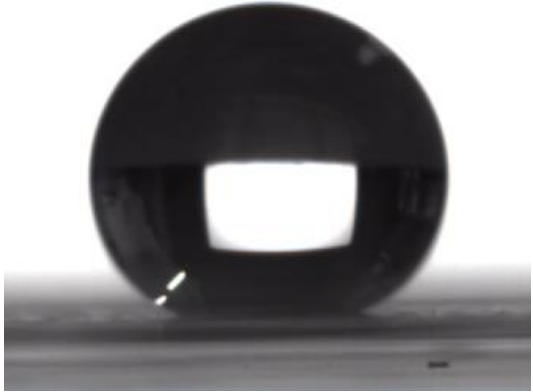

Contact angle of the upper plate (hydrophobic) in the droplet squeeze experiment. The contact angle is approximately 150°.

**Fig.SS5.**

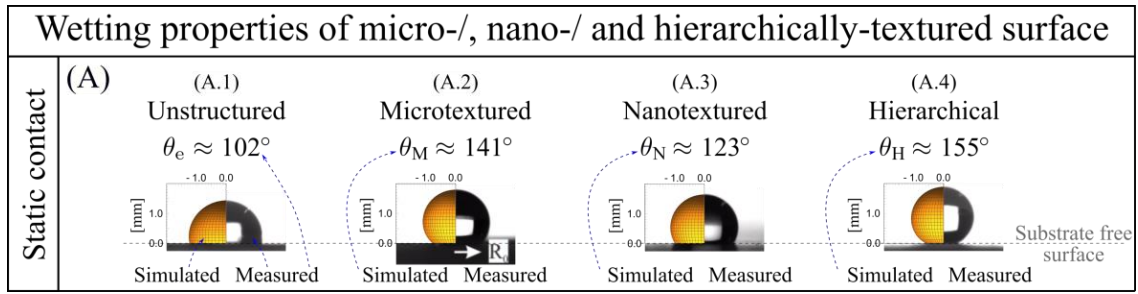

Experimental and simulation results of macroscopic static wetting properties (sessile droplet). (A) Measured and simulated droplet contact for the untextured (flat, A.1), microtextured (A.2), nanotextured (A.3) and hierarchical (A.4) surfaces.

**Fig.SS6.**

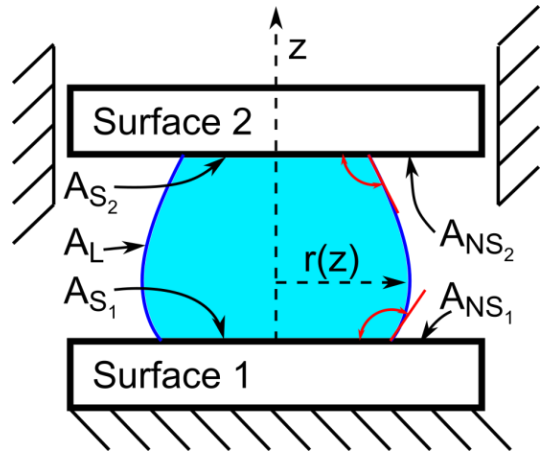

Schematic of the model describing a droplet squeezed among two dissimilar nominally flat surfaces.

**Fig.SS7.**

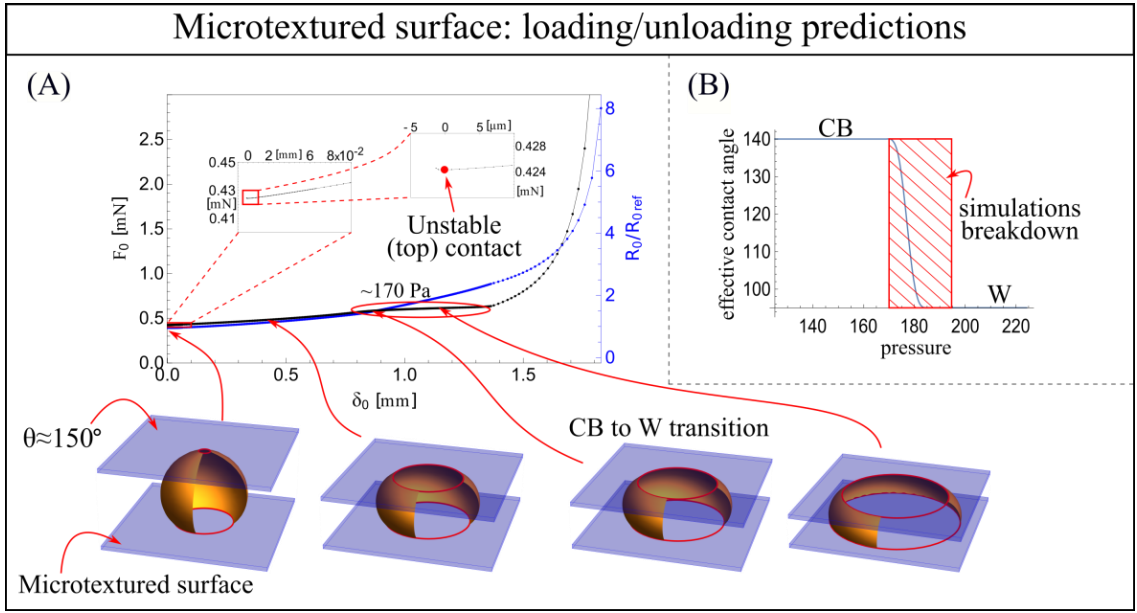

Simulation results: Force-displacement for the case where the droplet is squeezed between the microtextured surface (bottom plate) and a hydrophobic surface (top plate, see Fig.SS10). (A) Force (black line) and relative contact radius (blue line), predicted on the bottom plate, as a function of the top plate relative displacement. (B) Adopted relationship between apparent contact angle on the microstructure and pressure.

**Fig.SS8.**

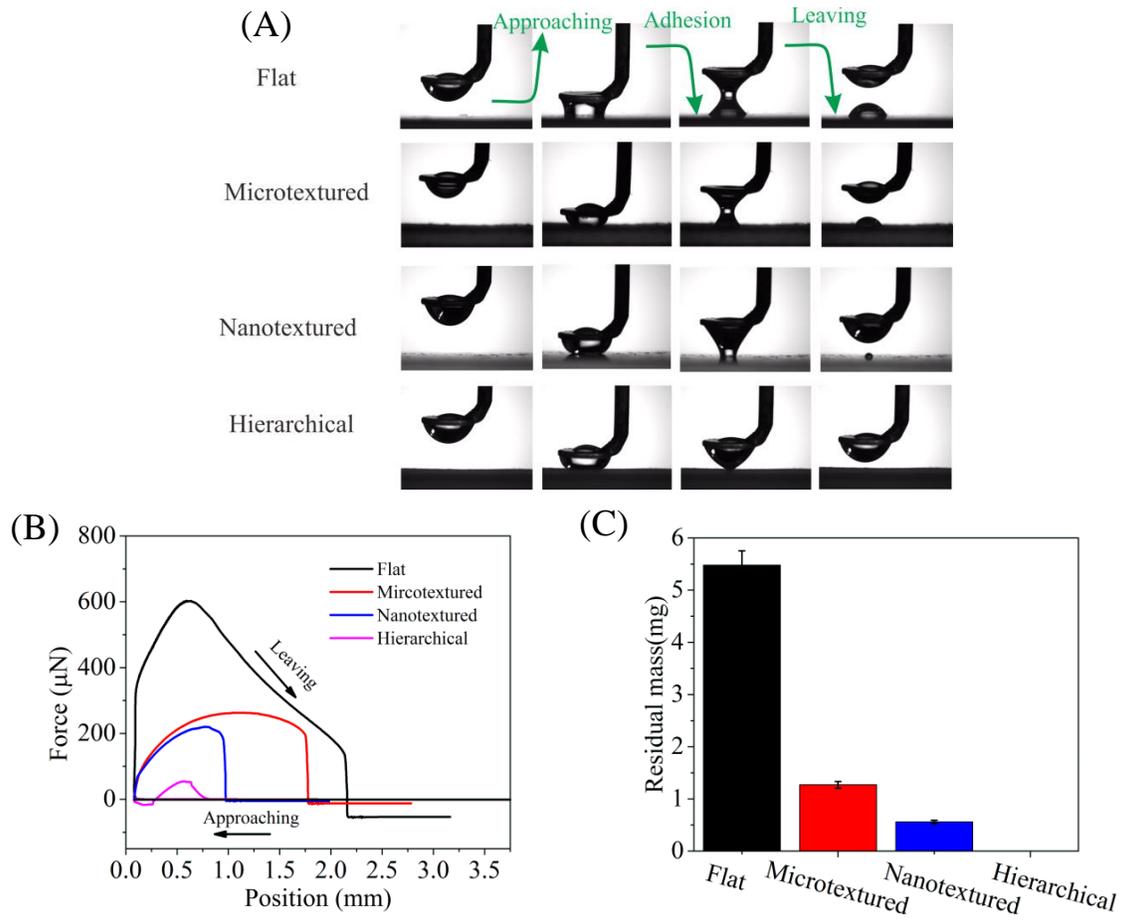

Wetting behaviour of the PP triboelectric layer when a water droplet is pressed against the different surfaces (flat, microtextured, nanotextured, hierarchical). (A) Photographic sequence (using a CCD camera) showing the loading and unloading stages of the adherence measuring process for the four different surfaces. (B) Force-distance curves of the water droplet (15 μL) approaching, contacting, leaving, and separating process with the PP triboelectric layer. (c) Residual mass of the water droplet after measurement.

**Fig.SS9.**

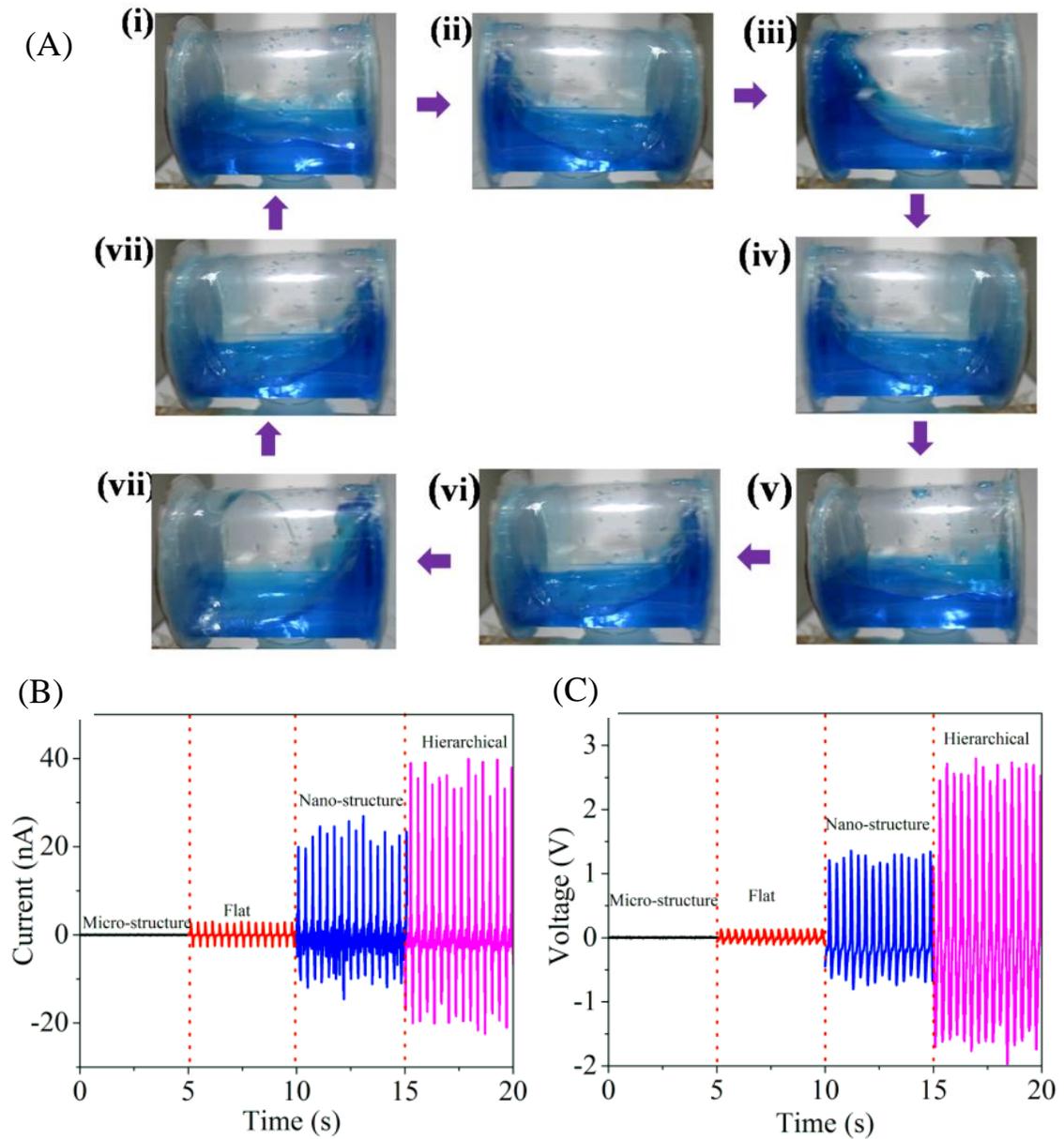

(A) Typical sloshing cycle adopted for the triboelectrical power generation of the water-based TENG, with water-to-cylinder volume ratio 0.3, vibration frequency 3 Hz, and vibration amplitude 5 mm. (B) and (C) Experimental results showing the steady-state triboelectrical outputs generated from the water-based TENG.

**Fig.SS10.**

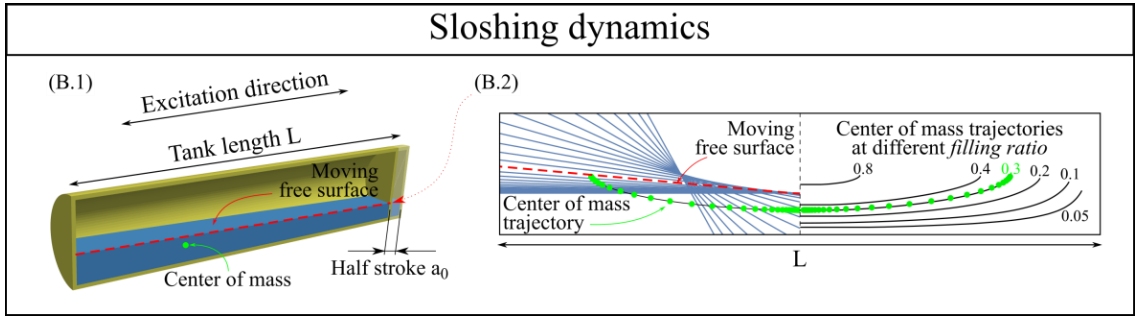

(A.1) Schematics of the water-based TENG, with indication of the excitation direction and parameters. (A.2) Left: Cross section of the tank showing the water free surface (red dashed line) and center of mass (green circle) trajectory during the sloshing. Right: center of mass trajectories at varying tank filling ratio. The green curve is for the sloshing condition which provides nearly the highest sloshing pressures, adopted through the following simulations and experiments.

**Fig.SS11.**

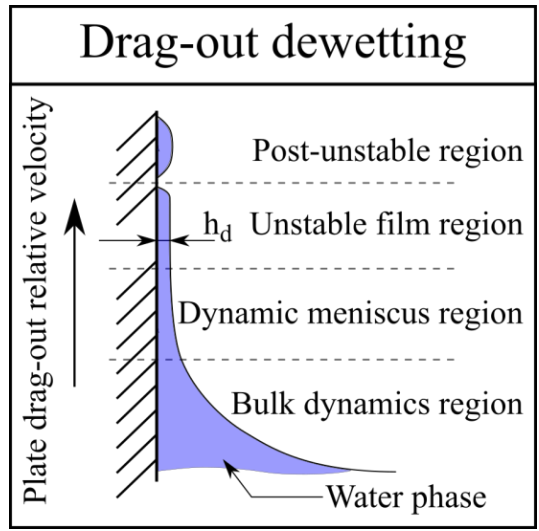

Schematics showing the fluid dynamics mechanisms involved in the drag-out dewetting process, with indication of the bulk region, the dynamic meniscus region, the unstable film region (if any), and the post-unstable region (if any). The existence of unstable and post-unstable ranges depends on the availability of a spontaneous film breakdown and external force-induced dewetting.

**Fig.SS12.**

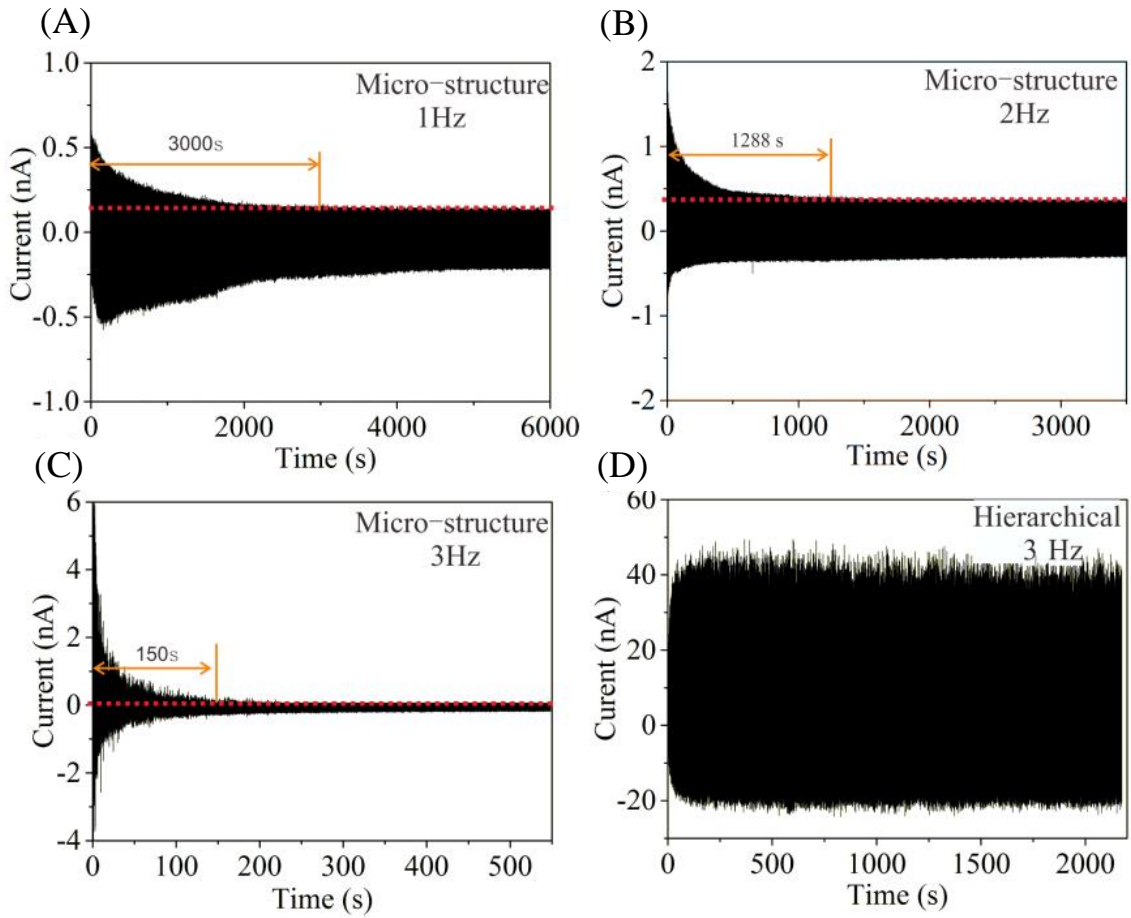

Measured transient output current for the TENG with microtextured PP surface at different frequency: (A) 1 Hz, (B) 2 Hz, (C) 3 Hz. (D) Measured transient output currents for the TENG with hierarchically structured PP surface at 3Hz.

**Fig.SS13.**

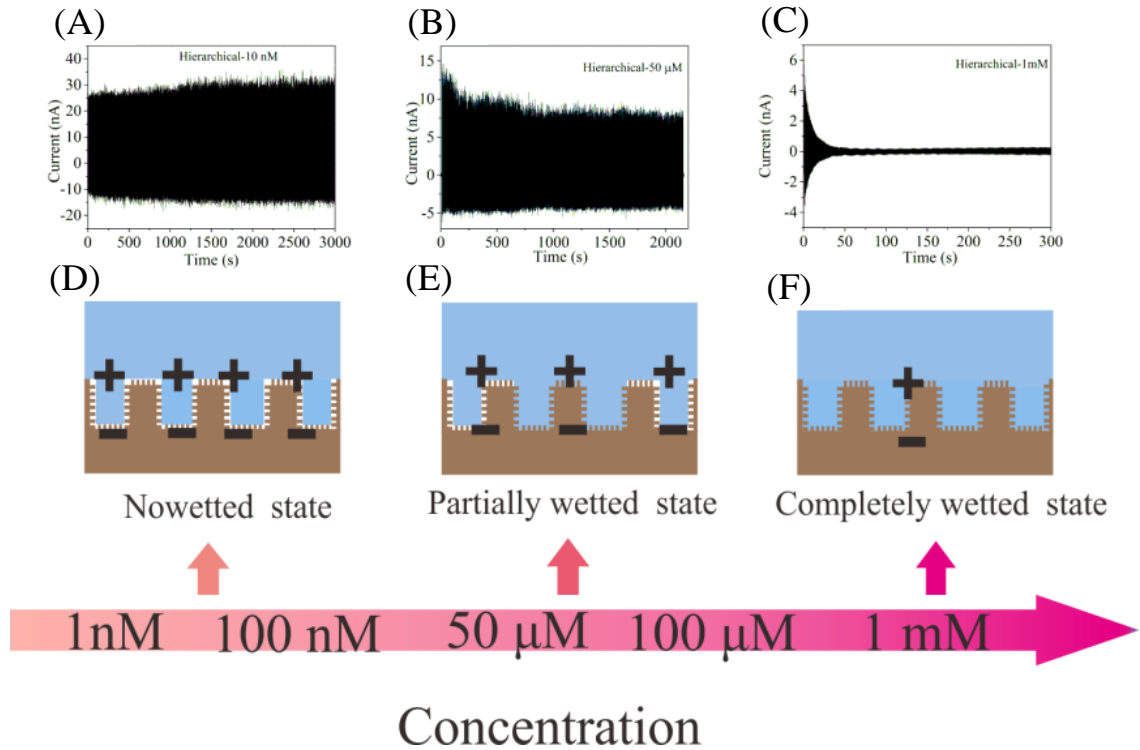

Full Wenzel transition with the increase of the concentration of surfactant. Measured transient tribocurrent of the water-based hierarchical TENG as a function of time for different surfactant concentration: (A) 10 nM, (B) 50 µM, and (C) 1 mM. Schematics of the wetting transitions occurring during surfactant injection: (D) nowetted state, (E) partially wetted state, and (F) full wetted state.

**Fig.SS14.**

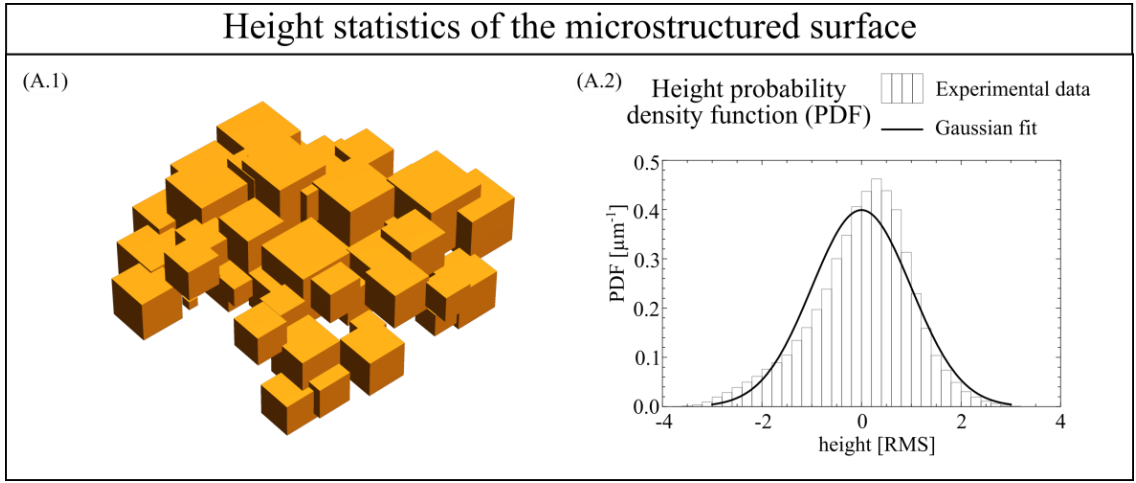

(A.1) Subset of microcubes of a generic microtextured virtual prototype. (A.2) Height probability density function of the microtextured surface roughness, skewed by the microcube geometries.

**Fig.SS15.**

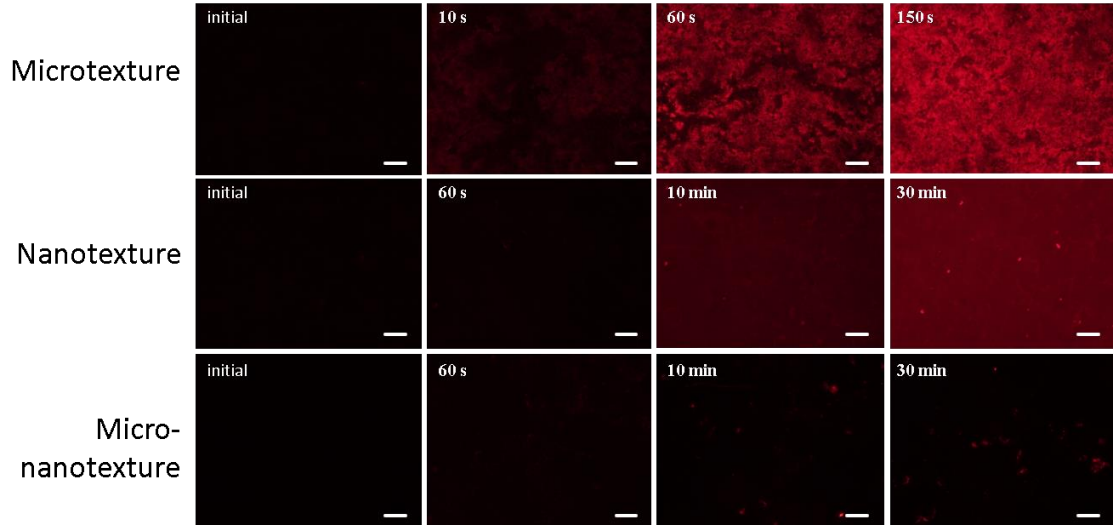

Epifluorence imaging of the residual wetting (irreversibly infiltrated fluid) on the textured PP surfaces at different sloshing times (3Hz sloshing frequency). RodhamineB is dissolved in water with concentration 0.01mg/ml. The bar corresponds to 200 µm.

**Fig.SS16.**

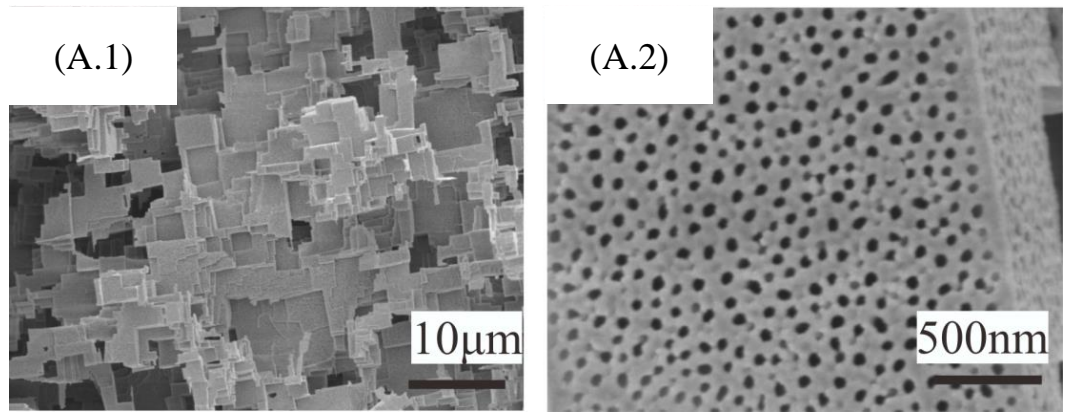

FESEM images of the hierarchical AAO templates with the low (A.1) and high (A.2) magnifications.

**SI References**

1. M. Cheng, *et al.*, Surface adhesive forces: a metric describing the drag-reducing effects of superhydrophobic coatings. *Small* **11**(14), 1665-1671 (2015).
2. H. Zou, *et al.*, Quantifying the triboelectric series. *Nat. Commun.* **10**(1), 1427 (2019).
3. D. Choi, *et al.*, Spontaneous electrical charging of droplets by conventional pipetting. *Sci. Rep.* **3**, 2037 (2013).
4. X. Zhang, *et al.*, Solid-liquid triboelectrification in smart U-tube for multifunctional sensors. *Nano Energy* **40**, 95-106 (2017).
5. R. A. Ibrahim, *et al.*, Recent advances in liquid sloshing dynamics. *Appl. Mech. Rev.* **54**(2), 133-199 (2001).
6. E. Rolley, C. Guthmann, R. Gombrowicz, V. Repain, Roughness of the Contact Line on a Disordered Substrate. *Phys. Rev. Lett.* **80**(13), 2865 (1998).
7. P. Volodin, A. Kondyurin, Dewetting of thin polymer film on rough substrate: I. Theory. *J. Phys. D: Appl. Phys.* **41**, 065306 (2008).